\documentclass[%
 reprint,
superscriptaddress,
footinbib,
 amsmath,amssymb,
 aps,
]{revtex4-1}
\usepackage{graphicx}
\usepackage{epstopdf}
\usepackage{caption}
\usepackage{subcaption} 
\usepackage{stackengine} 
\usepackage{float} 
\usepackage{dcolumn}
\usepackage{bm}
\usepackage{hyperref}

\begin{document}

\preprint{APS/123-QED}

\title{Power-law scaling in granular rheology across flow geometries}

\author{Seongmin Kim}
\affiliation{Harvard John A. Paulson School of Engineering and Applied Sciences, Harvard University, Cambridge, Massachusetts 02138, USA}
    
\author{Ken Kamrin}
\affiliation{Department of Mechanical Engineering, MIT, Cambridge, Massachusetts 02139, USA}%

\date{\today}
             
\begin{abstract}
Based on discrete element method simulations, we propose a new form of the constitution equation for granular flows independent of packing fraction. Rescaling the stress ratio $\mu$ by a power of dimensionless temperature $\Theta$ makes the data from a wide set of flow geometries collapse to a master curve depending only on the inertial number $I$. The basic power-law structure appears robust to varying particle properties (e.g. surface friction) in both 2D and 3D systems. We show how this rheology fits and extends frameworks such as kinetic theory and the Nonlocal Granular Fluidity model.

\end{abstract}



\maketitle



Granular materials exhibit complex mechanical behaviors: depending on the situation, they can either sustain loads like solids or flow like fluids. Diverse attempts have been made to build a continuum model for granular flows.
The $\mu(I)$ rheology \cite{GDR2004,daCruz2005,Jop2006}, a phenomenological model, suggests a one-to-one relation between two local dimensionless variables, the shear-to-normal stress ratio $\mu=\tau/P$ and the inertial number $I\equiv \dot{\gamma}/\sqrt{P/\rho_s d^2}$ for 3D spheres and $I\equiv \dot{\gamma}/\sqrt{P/m}$ for 2D disks 
where $\tau$ is the shear stress, $P$ is the pressure, $\dot{\gamma}$ is the shear rate, $\rho_s$ is the particle density, and $d$ and $m$ are respectively the mean particle diameter and mass. In this model, the shear rate vanishes if $\mu$ is smaller than a bulk friction coefficient $\mu_s$, and $I$ monotonically increases as $\mu$ increases for $\mu>\mu_s$.

However, this one-to-one relation between $\mu$ and $I$ loses accuracy outside of homogeneous shear flows. In general, nonlocal phenomena deviate flow from the $\mu(I)$ rheology~\cite{Pouliquen1999, Komatsu2001, GDR2004, Jop2007, Koval2009, Nichol2010, Reddy2011, Wandersman2014, Martinez2016, Tang2018}. To reconcile this deviation, nonlocal models such as the Nonlocal Granular Fluidity (NGF) model ~\cite{Kamrin2012, Kamrin2014, Kamrin2015, Henann2013} have been proposed. Inspired by a nonlocal model for emulsion flows~\cite{Goyon2008, Bocquet2009}, the NGF model assumes that a scalar ``fluidity'' field $g$ enters the flow rule through $\dot{\gamma}=g\mu$ and follows a phenomenological reaction-diffusion differential equation where the fluidity is generated by shearing and diffuses in space. Recently, through 3D discrete element method (DEM) simulations, Zhang and Kamrin~\cite{Zhang2017} have found that the fluidity field can be represented kinematically by the velocity fluctuations $\delta v$ and the packing fraction $\phi$: $g={\dot{\gamma}}/{\mu}=F(\phi){\delta v}/d$. Since $g$, the single evolving state field of the NGF model, appears to arise from \emph{two} kinematically observable state fields, we are motivated to seek further possible reductions. 

Interestingly, kinetic theory, which mathematically derives the constitutive equations using the Chapman-Enskog method, predicts a similar  relation between $\mu$, $\dot{\gamma}$, $\delta v$, and $\phi$. Introducing a granular temperature $T\equiv \delta v^2/D$ where $D$ is the spatial dimensions, kinetic theory predicts the pressure as $P=\rho_s F_1(\phi) T$ and the shear stress as $\tau=\rho_s F_2(\phi)\sqrt{T}\dot{\gamma}\,d$ where $F_1(\phi)$ and $F_2(\phi)$ depend on the radial distribution function~\cite{Jenkins1983, Lun1984, Garzo1999, Jenkins2010}. Thus, kinetic theory asserts $\mu=\left({F_2(\phi)}/{F_1(\phi)}\right){\dot{\gamma}d}/{\sqrt{T}}$ which becomes identical to Zhang's relation if ${F_1(\phi)}/{F_2(\phi)}=\sqrt{3}F(\phi)$. According to kinetic theory, since $\phi$ can be substituted by a function of dimensionless granular temperature $\Theta \equiv {\rho_s T}/{P}$, $\mu/I$ should be expressible as a function of $\Theta$. Although the assumptions of standard kinetic theory become less accurate near the jammed state, we are intrigued to consider whether some generic $\mu(I, \Theta)$ relation continues to exists into the dense regime, effectively removing rheological dependence on $\phi$. The notion of expanding the $\mu(I)$ model by dimensionless temperature has also been considered in \cite{Gaume2011}, which we shall discuss later.

To explore a potential $\mu(I, \Theta)$ relation, we take a hint from the power-law dependencies of thermodynamic quantities in many complex systems which exhibit continuous phase transitions. Near the critical temperature $T_c$ where the microscopic entities are highly correlated, the macroscopic fields follow scaling forms characterized by a power function of the reduced temperature $(T-T_c)/T_c$ ~\cite{Kardar2007}. Although granular systems are athermal, the velocity fluctuations created by shearing may act like the temperature. Moreover, previous studies have observed more correlated motion of grains as a granular material approaches the jammed state~\cite{Radjai2002, staron2002preavalanche, Pouliquen2004, Silbert2005}. It is thus natural to suspect power-law scaling in a $\mu(I,\Theta)$ relation as a possible unifying principle in granular rheology.

Inspired by critical scaling, in this Letter we show that rescaling $\mu$ by a simple power of $\Theta$  collapses data from many DEM strongly onto a master curve that depends only on $I$. In doing so, we identify and validate a general relation of the form $\mu(I,\Theta)$ that holds across geometries and flow regimes.


\begin{figure}
        \includegraphics[width=0.46\textwidth]{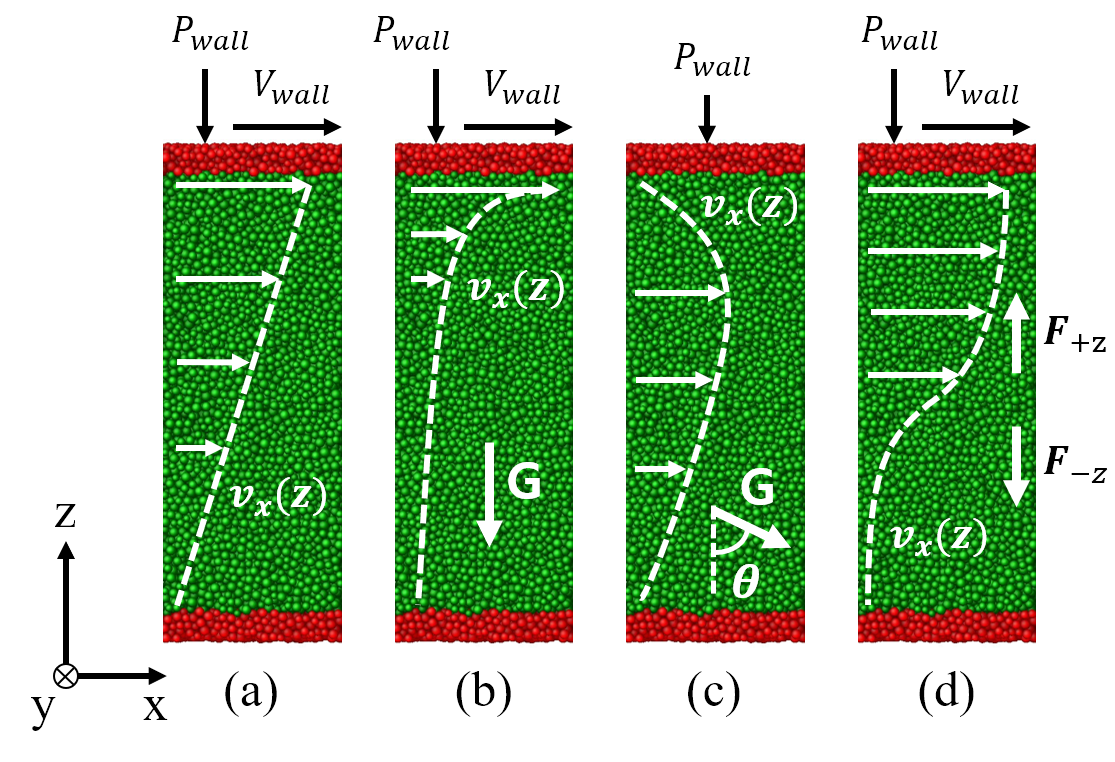}
        \captionsetup{justification=raggedright}
        \caption{ Planar shear geometries tested: (a) simple shear, (b) shear with gravity, (c) chute flows ($\theta={60}^{\circ}$ and $90^{\circ}$), and (d) concave flows. The dashed lines are schematic velocity profiles.}\captionsetup{justification=justified}
        \label{fig:1}
\end{figure}

\begin{figure*}
    \centering
    \begin{subfigure}[t]{0.227\textwidth}
        \centering
        \phantomcaption
        \stackinset{l}{0.87\textwidth}{b}{0.18\textwidth}
        {(\thesubfigure)}
        {\includegraphics[width=\textwidth]{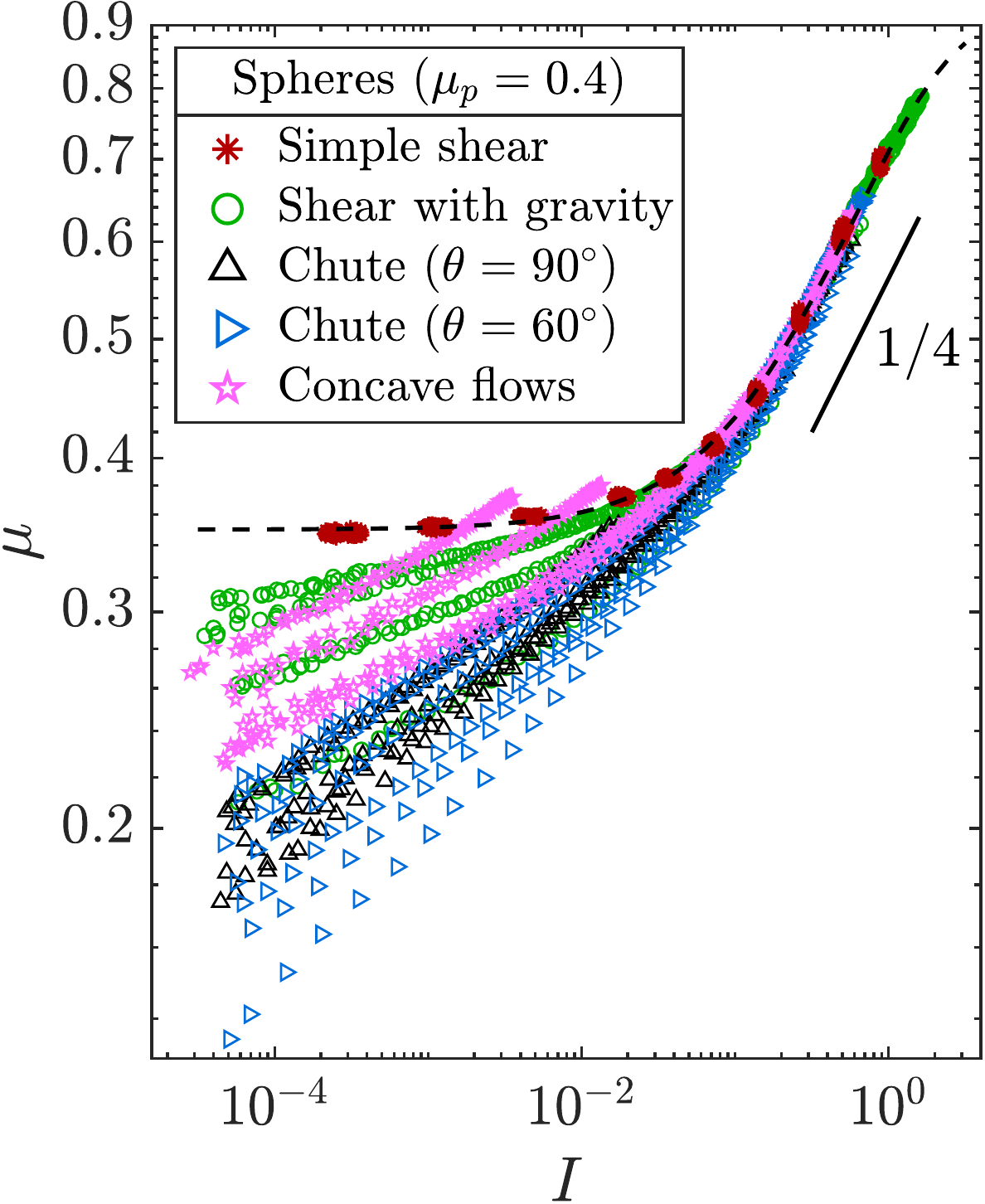}}
        \label{fig:2a}
    \end{subfigure}
    \quad
    \begin{subfigure}[t]{0.232\textwidth}  
        \centering
        \phantomcaption
        \stackinset{l}{0.87\textwidth}{b}{0.18\textwidth}
        {(\thesubfigure)}
        {\includegraphics[width=\textwidth]{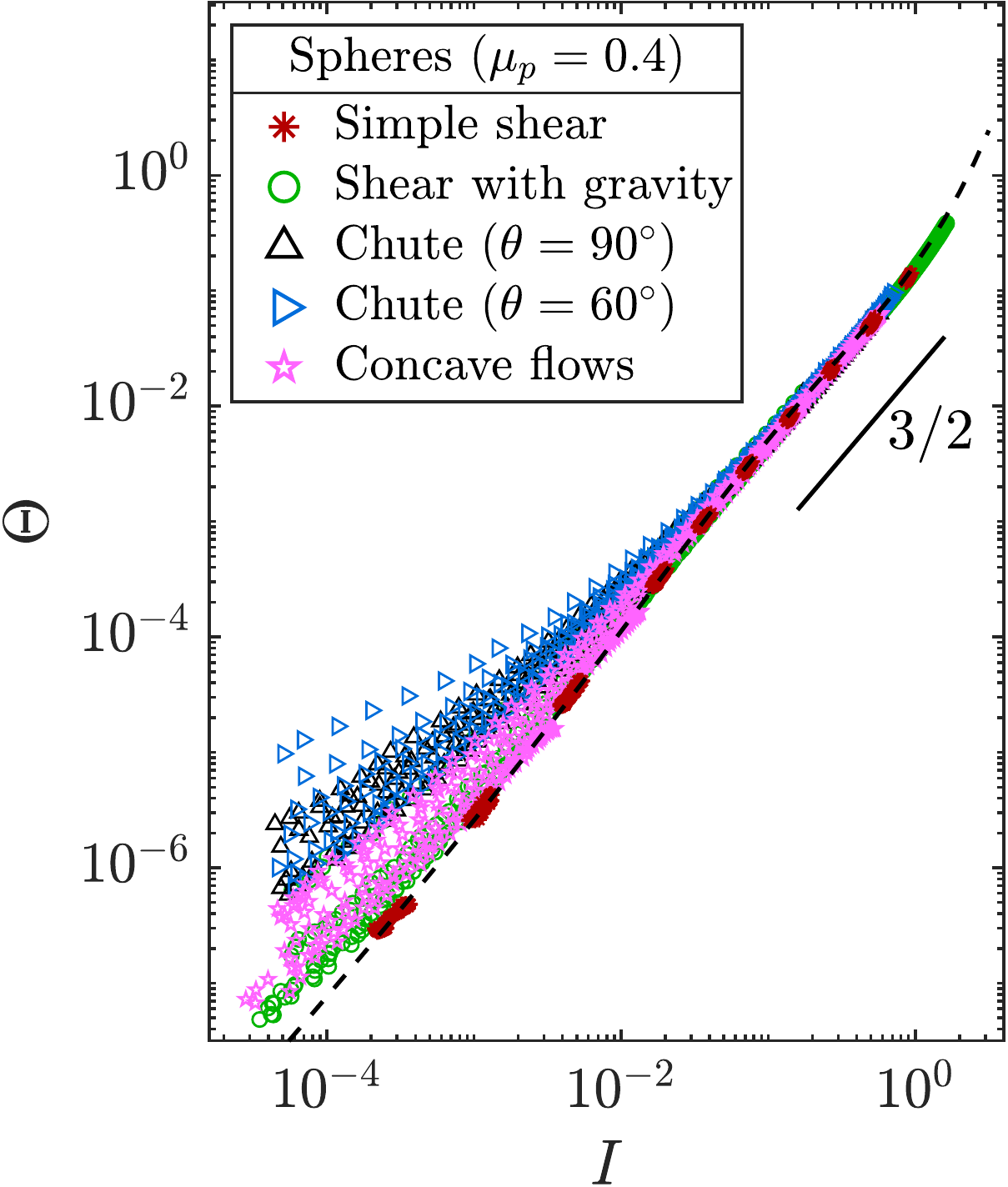}}
        \label{fig:2b}
    \end{subfigure}
    \quad
    \begin{subfigure}[t]{0.232\textwidth}  
        \centering 
        \phantomcaption
        \stackinset{l}{0.87\textwidth}{b}{0.18\textwidth}
        {(\thesubfigure)}
        {\includegraphics[width=\textwidth]{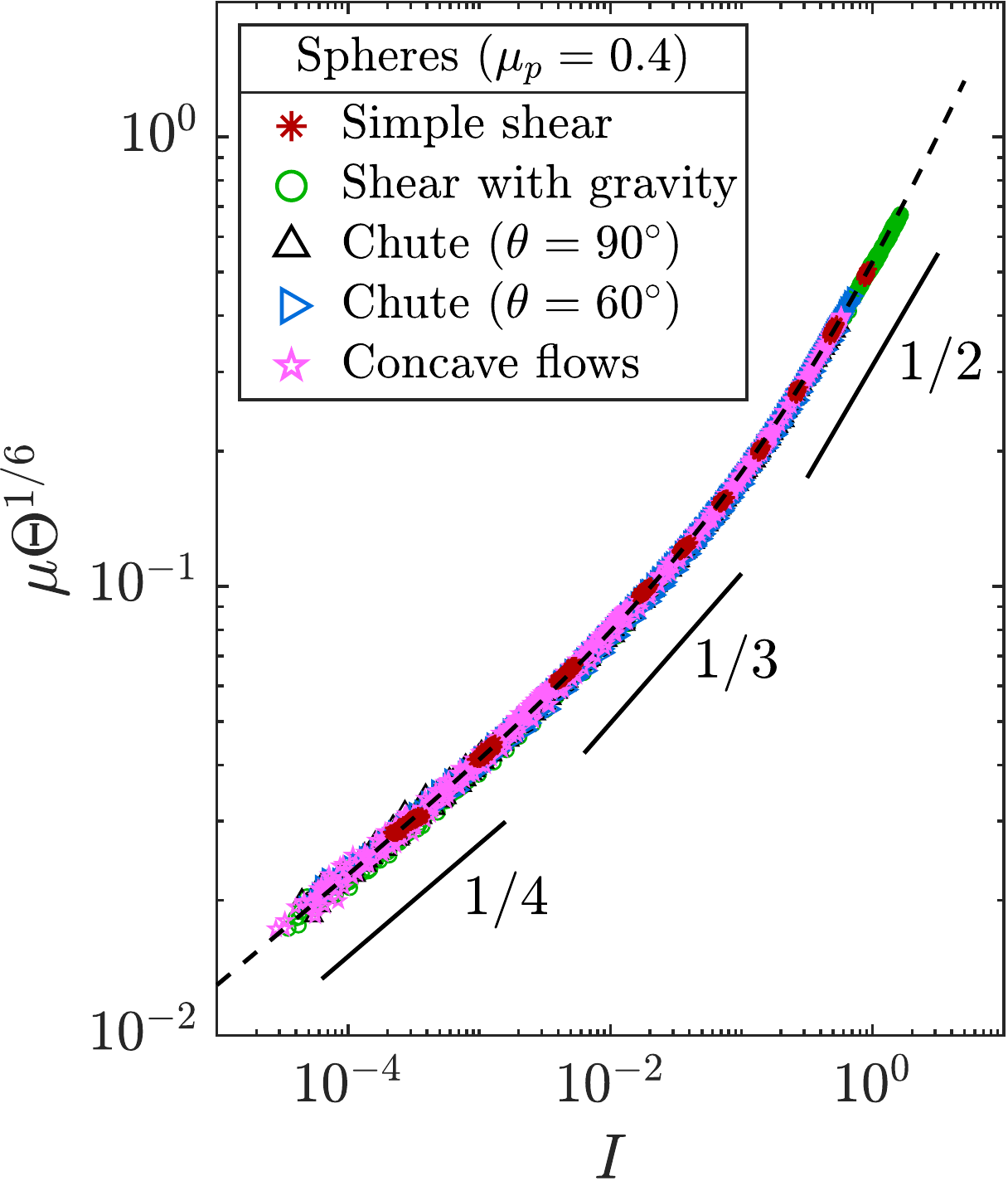}}
        \label{fig:2c}
    \end{subfigure}
    \quad
    \begin{subfigure}[t]{0.232\textwidth}  
        \centering 
        \phantomcaption
        \stackinset{l}{0.87\textwidth}{b}{0.18\textwidth}
        {(\thesubfigure)}
        {\includegraphics[width=\textwidth]{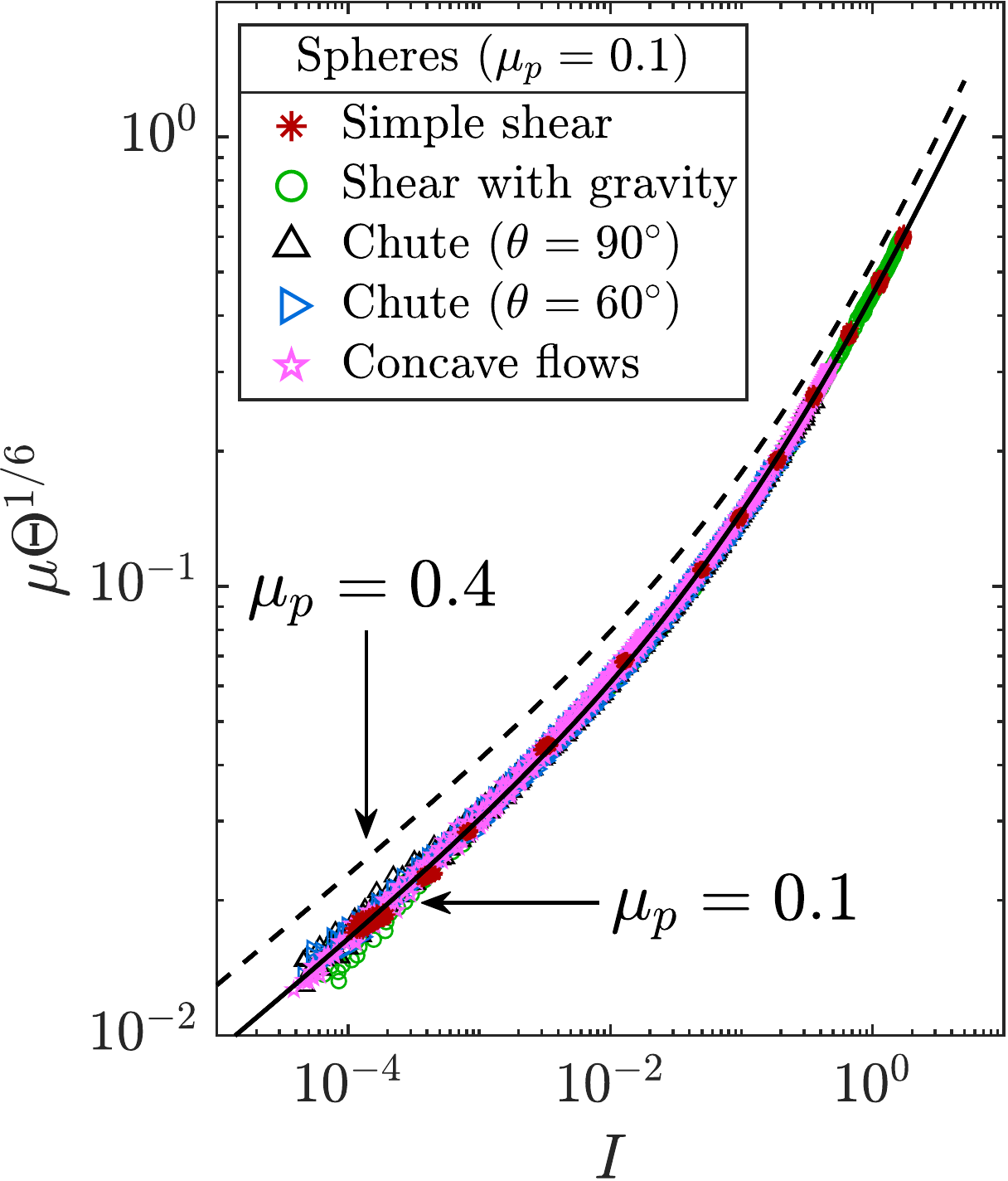}}
        \label{fig:2d}
    \end{subfigure}
    \\    
    \begin{subfigure}[t]{0.227\textwidth}   
        \centering 
        \phantomcaption
        \stackinset{l}{0.87\textwidth}{b}{0.18\textwidth}
        {(\thesubfigure)}
        {\includegraphics[width=\textwidth]{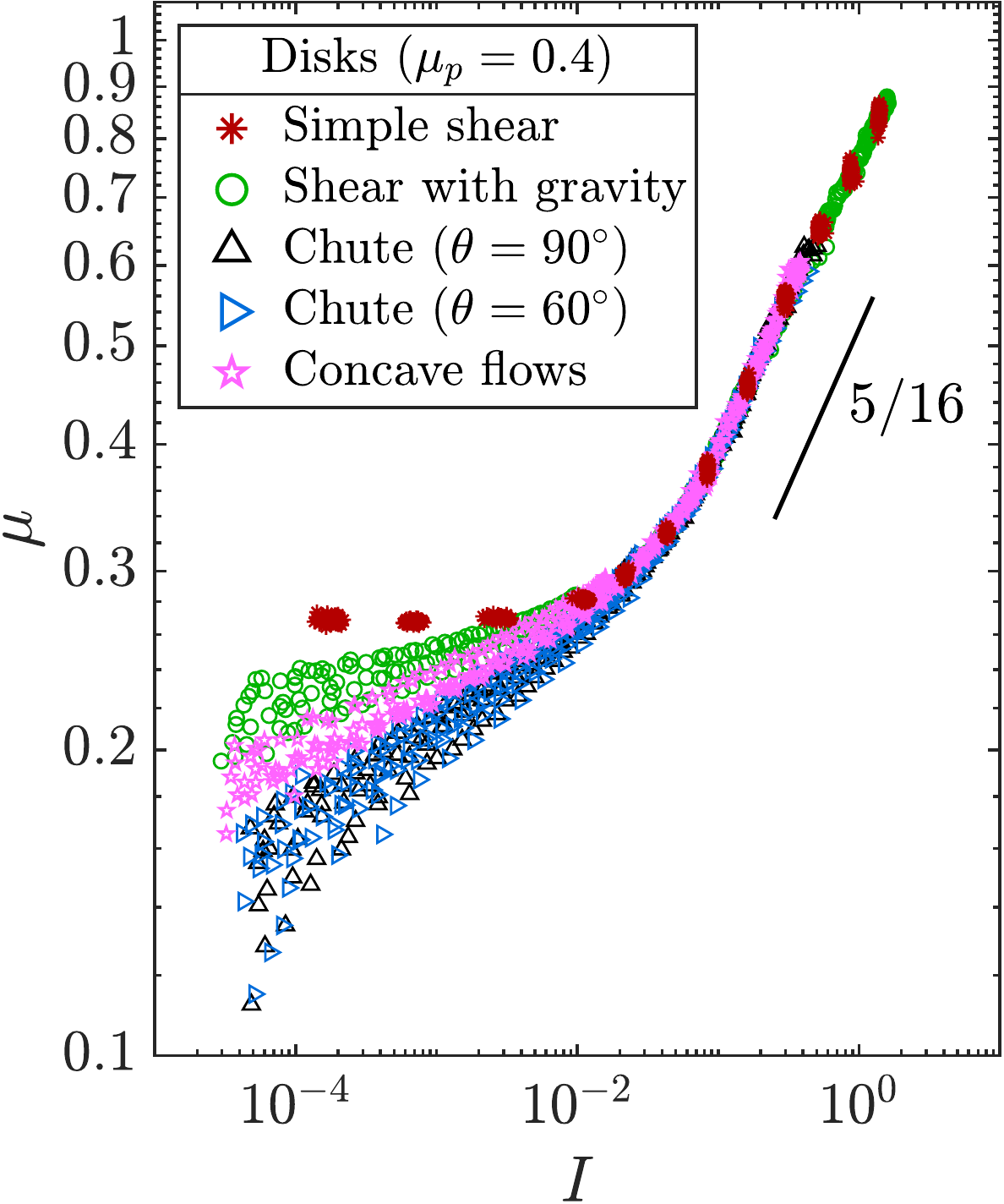}}
        \label{fig:2e}
    \end{subfigure}
    \quad
    \begin{subfigure}[t]{0.232\textwidth}   
        \centering 
        \phantomcaption
        \stackinset{l}{0.87\textwidth}{b}{0.18\textwidth}
        {(\thesubfigure)}
        {\includegraphics[width=\textwidth]{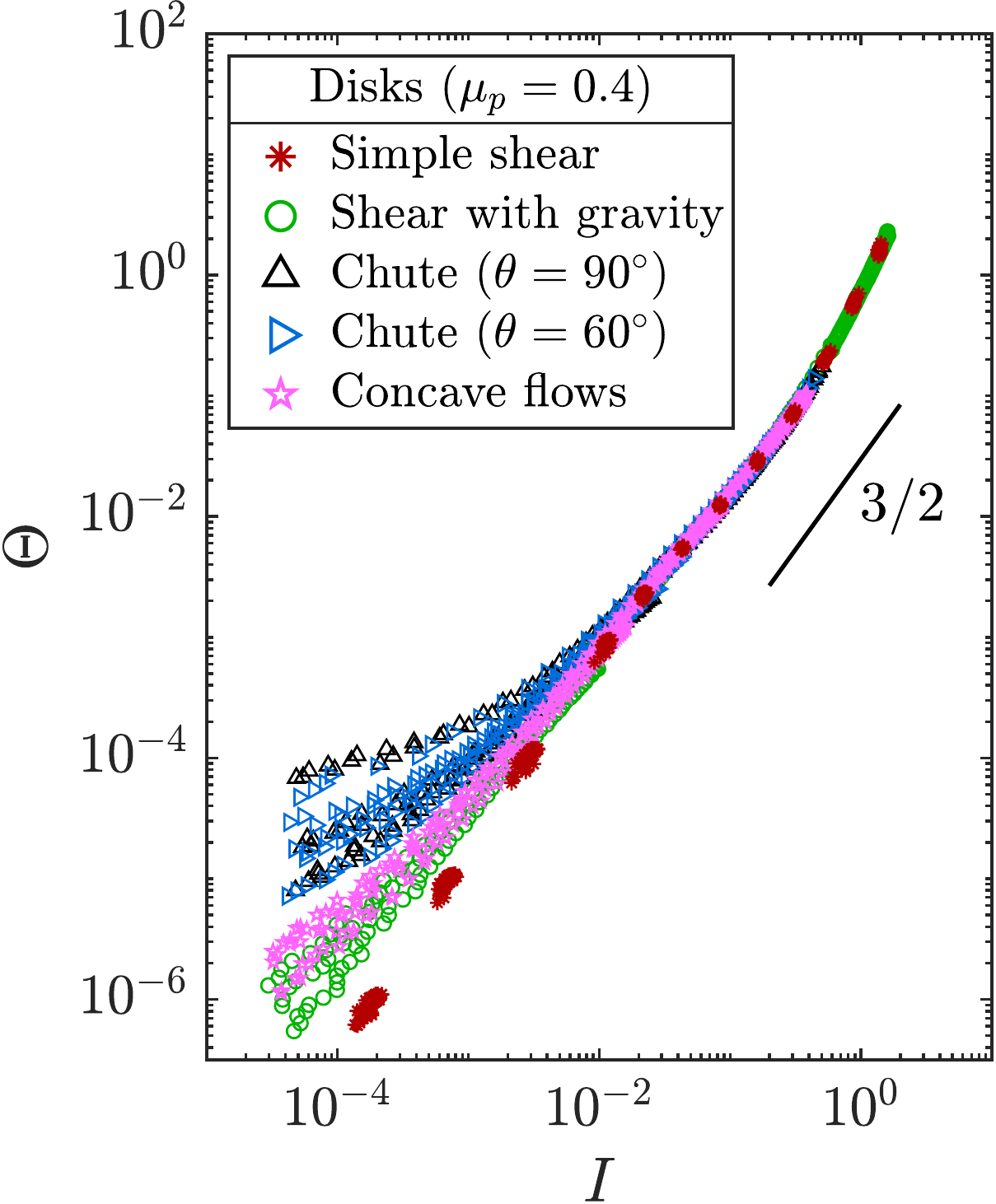}}
        \label{fig:2f}
    \end{subfigure}
    \quad
    \begin{subfigure}[t]{0.232\textwidth}  
        \centering 
        \phantomcaption
        \stackinset{l}{0.87\textwidth}{b}{0.18\textwidth}
        {(\thesubfigure)}
        {\includegraphics[width=\textwidth]{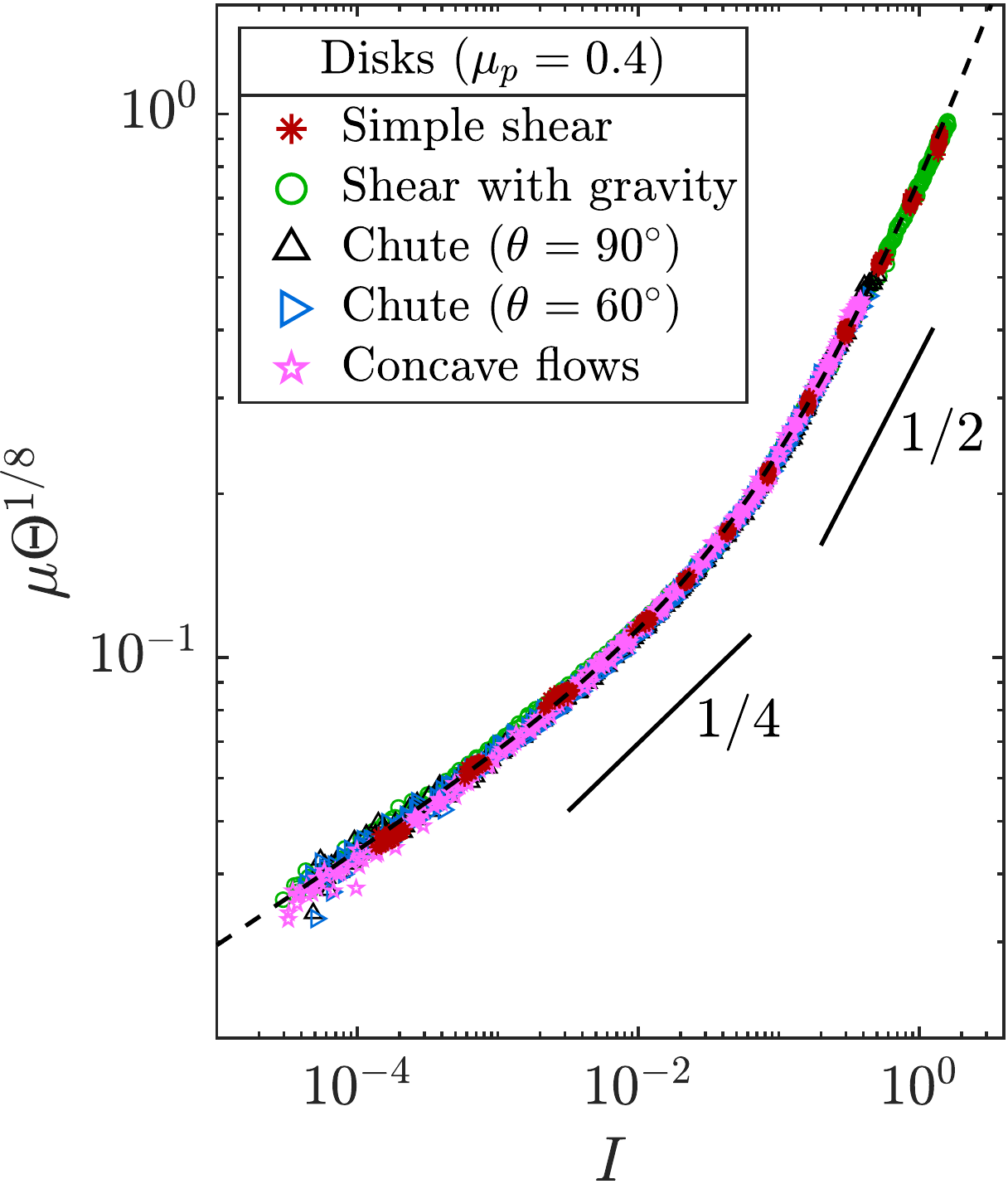}}
        \label{fig:2g}
    \end{subfigure}
    \quad
    \begin{subfigure}[t]{0.232\textwidth}  
        \centering 
        \phantomcaption
        \stackinset{l}{0.87\textwidth}{b}{0.18\textwidth}
        {(\thesubfigure)}
        {\includegraphics[width=\textwidth]{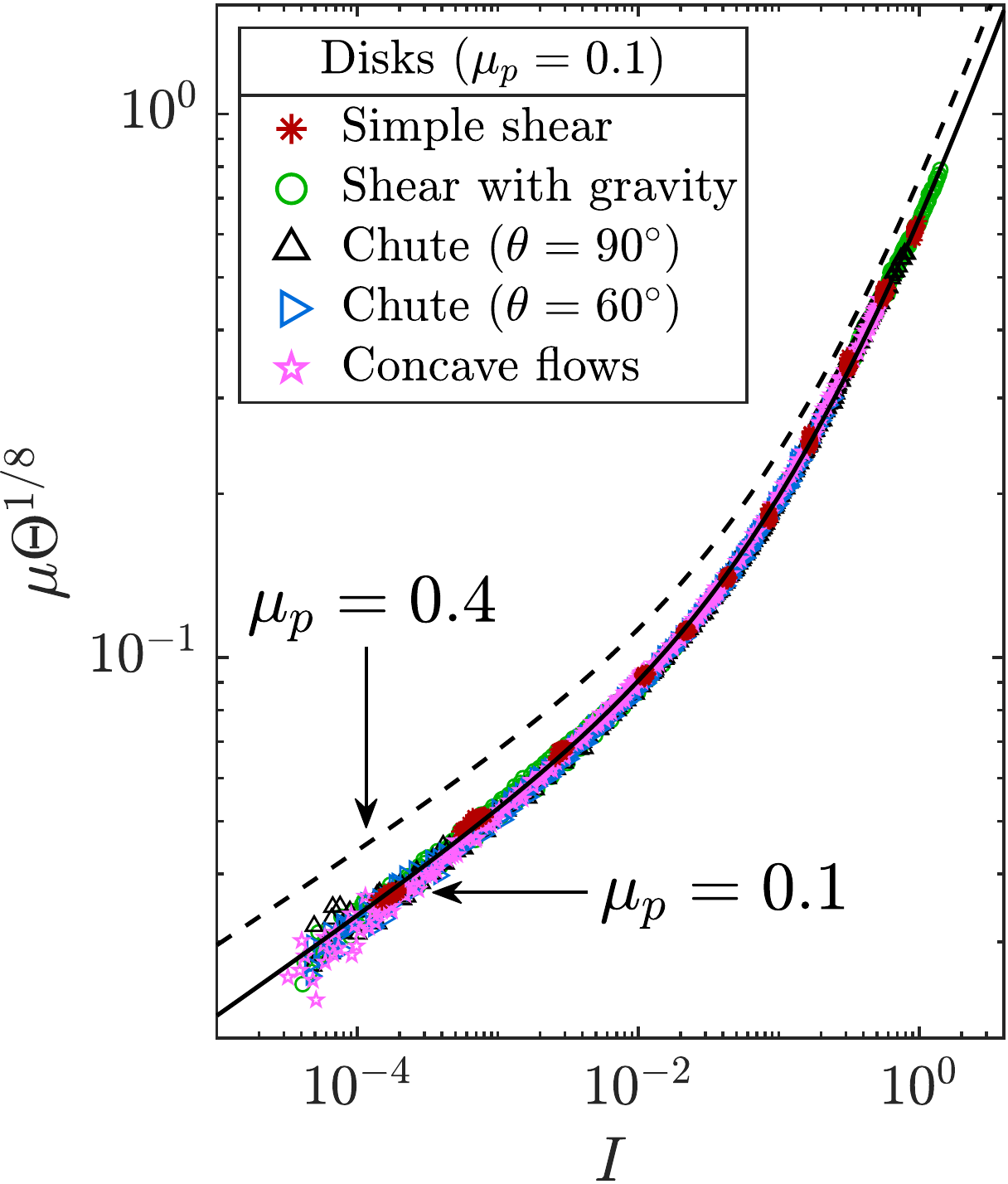}}
        \label{fig:2h}
    \end{subfigure}
     \captionsetup{justification=raggedright, singlelinecheck=false}
    \caption{
    Relations between $\mu$, $I$, and $\Theta$ in various planar shear configurations. Non-collapse of $\mu$ vs $I$  in 3D (\protect\subref{fig:2a}) and 2D (\protect\subref{fig:2e}), and similar non-collapse of $\Theta$ vs $I$ (\protect\subref{fig:2b}, \protect\subref{fig:2f}). The dashed lines in (\protect\subref{fig:2a}) and (\protect\subref{fig:2b}) are $\mu_{loc}(I)$ and $\Theta_{loc}(I)$ in Eq.~\eqref{eq:3} respectively.
    Multiplying $\mu$ by $\Theta^p$ makes the scattered points collapse into a master curve $f(I)$ (dashed trend lines for $\mu_p=0.4$ and solid trend lines for $\mu_p=0.1$). In 3D, $p=1/6$ (\protect\subref{fig:2c},~\protect\subref{fig:2d}) and, in 2D, $p=1/8$ (\protect\subref{fig:2g},~\protect\subref{fig:2h}). The surface friction does not change the exponent but changes the master curve (\protect\subref{fig:2d},~\protect\subref{fig:2h}).}
    \label{fig:2}
    
\end{figure*}

We use LAMMPS to simulate granular flows of 3D spheres and 2D disks. The average diameter and the density of particles are denoted as $d$ and $\rho_s$ which gives the characteristic mass $m=\rho_s\pi d^3/6$ in 3D and $m=\rho_s\pi d^2/4$ in 2D. To prevent crystallization, we set the diameter of each particle to be uniformly distributed from $0.8d$ to $1.2d$. For the contact forces, we use the standard spring-dashpot model with the Coulomb friction as in previous studies~\cite{Cundall1979, daCruz2005, Koval2009, Zhang2017, Kamrin2012, Kamrin2014, Liu2018}.
In order to simulate hard particles, we choose the normal elastic constant high enough to keep the average overlapping distance smaller than $10^{-5}d$. The tangential elastic constant is set to be $2/7$ of the normal one. The damping coefficient is chosen to make the restitution coefficient to be $0.24$.

We perform simulations on planar shear flows with diverse body forces and boundary conditions per Fig.~\ref{fig:1}. Simple shear flows (Fig.~\ref{fig:1}a) generate the $\mu(I)$ rheology while shear flows with gravity (Fig.~\ref{fig:1}b), flows in a vertical chute (Fig.~\ref{fig:1}c with $\theta=90^\circ$), flows in a tilted chute (Fig.~\ref{fig:1}c with $\theta=60^\circ$), and ``concave'' flows (Fig.~\ref{fig:1}d) exhibit nonlocality. Concave flows are so-named after the shape of the shearing profile, which arises from an outward external force ${\vec F_z}\propto(m/d)(z-z_0)\hat{z}$ for $z_0$ the midpoint of the system. The gravity $G$ is constant for each case.
The simulated domain is cuboid ($L_x=20d$ and $L_y=16d$; $L_\alpha$ is the system length in the $\alpha$-direction) for 3D systems and rectangular ($L_x=160d$) for 2D. The horizontal boundaries are periodic. We employ a widely used feedback scheme to assert top-wall pressure $P_{wall}$ ~\cite{daCruz2005, Koval2009, Kamrin2014, Zhang2017, Liu2018}. The horizontal wall velocity $V_{wall}$ is constant. We use different $P_{wall}$ and $V_{wall}$ combinations to generate varied flow profiles. In total, we ran 105 different simulations, spanning two surface friction coefficients ($\mu_p=0.4$ and $0.1$) and two grain shapes (3D spheres and 2D disks). The total number of particles in each simulation varies from around $6.7\times10^3$ to $2.0\times10^4$. See Supplemental Material \footnote{See Supplemental Material for more discussion on DEM and continuum simulation methods, additional DEM data and continuum solutions, and $f(I)$ fit functions.} for more details.

When steady state is reached, the averaged continuum fields are calculated by coarse-graining. Following previous studies~\cite{Koval2009, Kamrin2014, Liu2018}, we calculate the instantaneous velocity field by $\vec v(z_k,t) = \sum_{i}{A_{ki} \vec v_i(t)}/\sum_{i}{A_{ki}}$ where $v_i$ is the velocity of the $i$th particle and $A_{ki}$ is the cross-sectional area (length in 2D) between the $i$th particle and the plane of $z=z_k$. The interval of $z_k$ is kept less than $0.5d$.  We define the instantaneous granular temperature tensor as $\bm{T}(t) = \sum_{i} A_{ki} (\delta \vec{v}_i(t) \otimes \delta \vec{v}_i(t))/ \sum_{i} A_{ki} $ where $\delta\vec{v_i}(t)=\vec{v_i}(t)-\vec{v}(z_k, t)$. When we calculate the velocity fluctuations, we use the instantaneous velocity field as in ~\cite{Zhang2017}. The instantaneous stress is given by $\bm{\sigma}(r_k,t)=\bm{\sigma}^{K}(r_k,t)+ \sum_{i}{A_{ki} \bm{\sigma} _i (t)}/A $ where $\bm{\sigma}_i $ is the particle-wise stress from contacts, $A$ is the area of the horizontal plane ($L_x$ in 2D), and  $\bm{\sigma}^{K} = -\rho_s\phi\,\bm{T}$ is the kinetic stress~\cite{Weinhart2013}. The granular temperature is chosen as $T_{xx}={\delta v_x}^2$ because the diagonal components are slightly different each other possibly due to rigid-wall effects. Similarly, we choose $P$ as $-\sigma_{zz}$, $\tau$ as $\left|\sigma_{xz}\right|$, and $\dot{\gamma}$ as $\left|\partial_z v_x\right|$. All the fields are then averaged over time. For well-averaged steady flow data within a limited number of snapshots excluding wall effects, we cut off the data where total local shear is less than 1, $\phi<0.4$, or the distance from the walls is less than $3d$.

The relations between the coarse-grained fields are shown in Fig.~\ref{fig:2}. As many previous studies have observed, $\mu$ and $I$ are not one-to-one in inhomogeneous flows (Fig.~\ref{fig:2a} and Fig.~\ref{fig:2e}). Also, $\Theta$ is not determined only by $I$ (Fig.~\ref{fig:2b}, Fig.~\ref{fig:2f}). However, there is a certain trend. For a given $I$, smaller $\mu$ corresponds to larger $\Theta$ as if heating softens the material. In the spirit of the power-law scaling in continuous phase transitions, we have tried multiplying either $\mu$ or $I$ by a power of $\Theta$, which are the simplest cases, changing the exponent $p$ to achieve the best data collapse. Surprisingly, all the 3D sphere data with $\mu_p=0.4$ gathers to a single master curve when $\mu$ is multiplied by $\Theta^{1/6}$: $\mu\Theta^{1/6}=f(I)$ (Fig.~\ref{fig:2c}). Rescaling $I$ does not give a better data collapse than rescaling $\mu$. The same exponent $p=1/6$ also works for $\mu_p=0.1$ cases, but the data points collapse to a lower master curve (Fig.~\ref{fig:2d}). Rescaling $\mu$ with a power of $\Theta$ also produces a well-collapsed master curve for disks, but the best exponent $p$ is about $1/8$ for both $\mu_p=0.1$ and $0.4$ cases (Fig.~\ref{fig:2g} and Fig.~\ref{fig:2h}). Therefore, we propose that for hard particles systems, 
\begin{equation} \label{eq:1}
    \mu\Theta^p = f(I)
\end{equation}
where $p$ depends on the spatial dimensions and $f(I)$ depends as well on particle information. See the Supplemental Material for the fitting functions in Fig.~\ref{fig:2}.


\begin{figure}
    \centering
    \begin{subfigure}[t]{0.23\textwidth}  
        \centering 
        \phantomcaption
        \stackinset{l}{-0.02\textwidth}{b}{0.91\textwidth}
        {(\thesubfigure)}
        {\includegraphics[width=\textwidth]{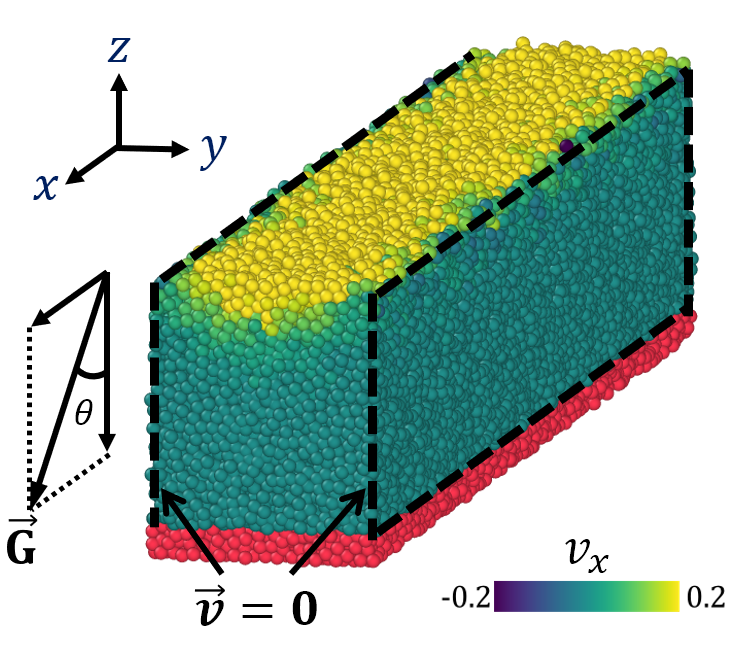}}
        \label{fig:3a}
    \end{subfigure}
    \quad
    \begin{subfigure}[t]{0.21\textwidth}  
        \centering 
        \phantomcaption
        \stackinset{l}{-0.02\textwidth}{b}{1.0\textwidth}
        {(\thesubfigure)}
        {\includegraphics[width=\textwidth]{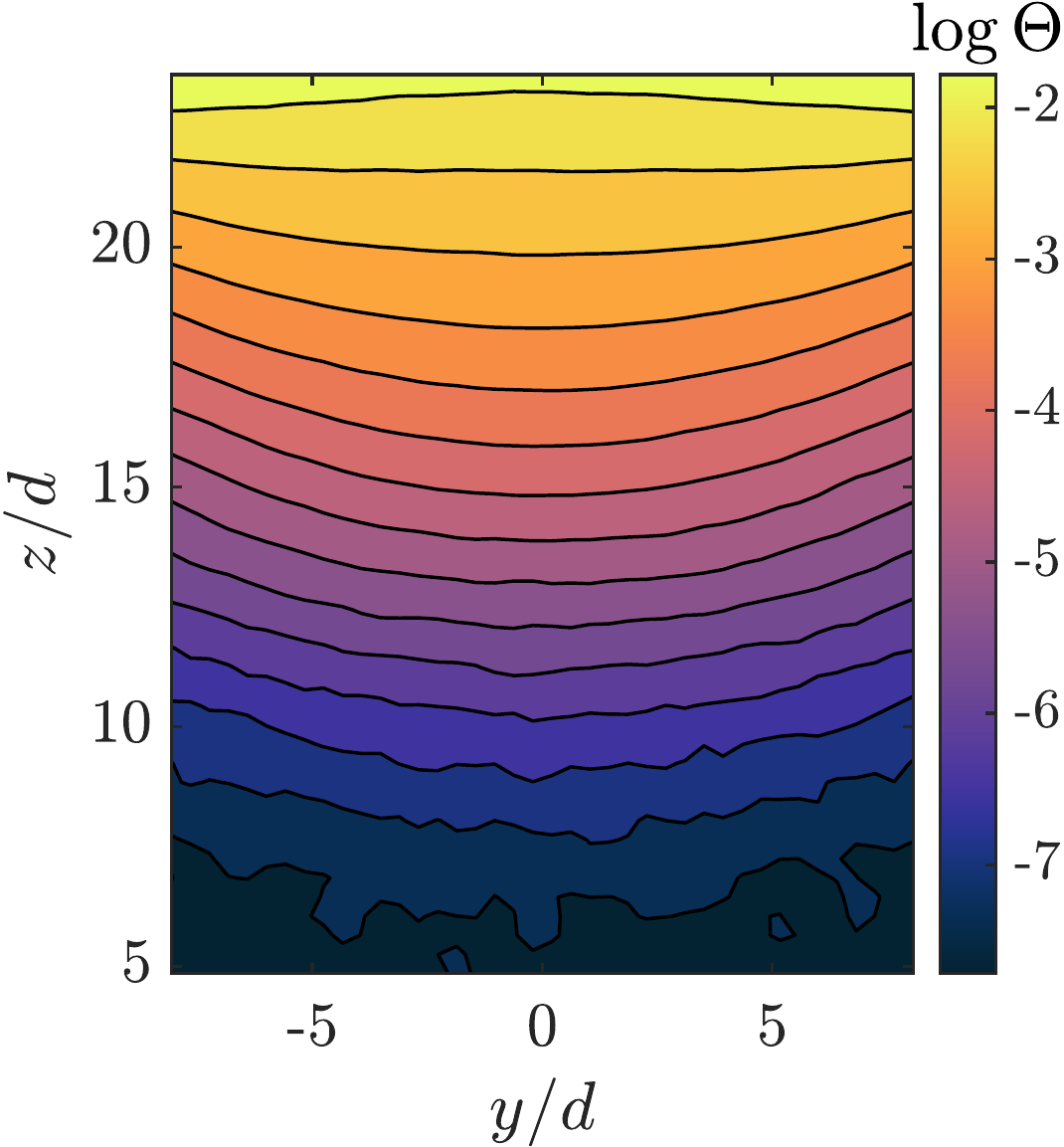}}
        \label{fig:3b}
    \end{subfigure}
    \\
    \begin{subfigure}[t]{0.22\textwidth}  
        \centering 
        \phantomcaption
        \stackinset{l}{-0.02\textwidth}{b}{1.0\textwidth}
        {(\thesubfigure)}
        {\includegraphics[width=\textwidth]{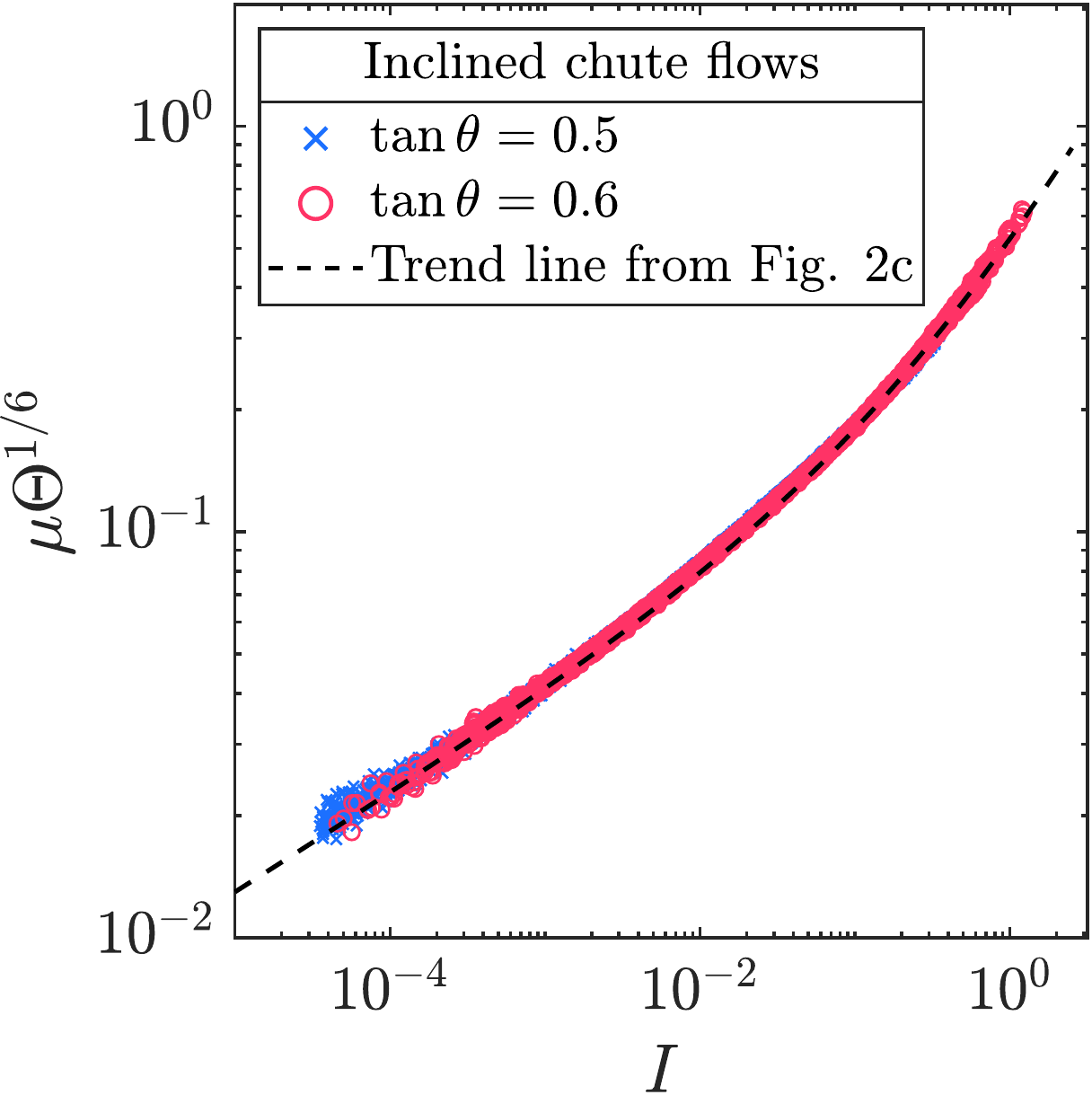}}
        \label{fig:3c}
    \end{subfigure}
    \quad
    \begin{subfigure}[t]{0.22\textwidth}  
        \centering 
        \phantomcaption
        \stackinset{l}{-0.03\textwidth}{b}{1.0\textwidth}
        {(\thesubfigure)}
        {\includegraphics[width=\textwidth]{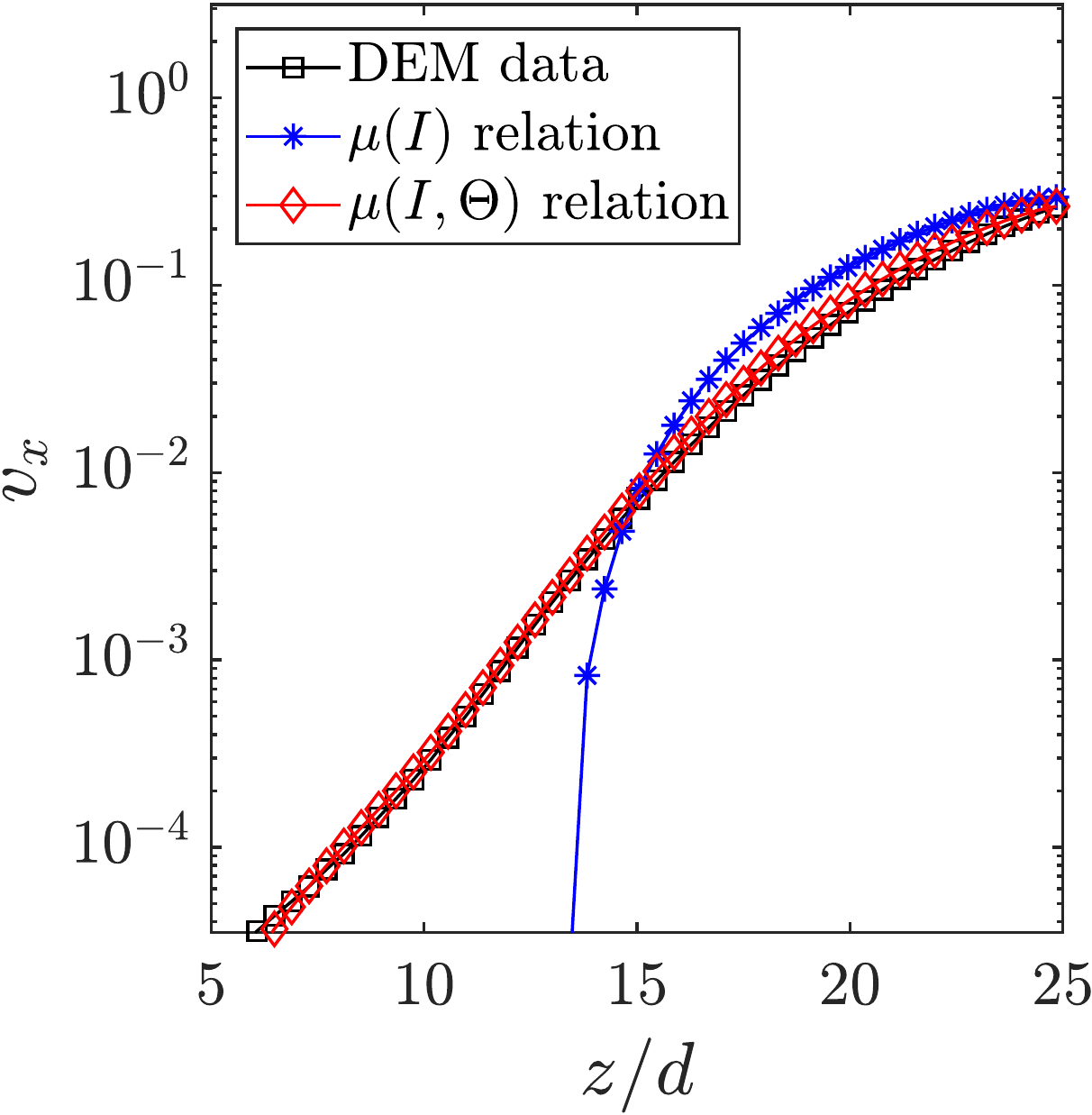}}
        \label{fig:3d}
    \end{subfigure}
    \\
    \begin{subfigure}[t]{0.22\textwidth}  
        \centering 
        \phantomcaption
        \stackinset{l}{-0.02\textwidth}{b}{1.0\textwidth}
        {(\thesubfigure)}
        {\includegraphics[width=\textwidth]{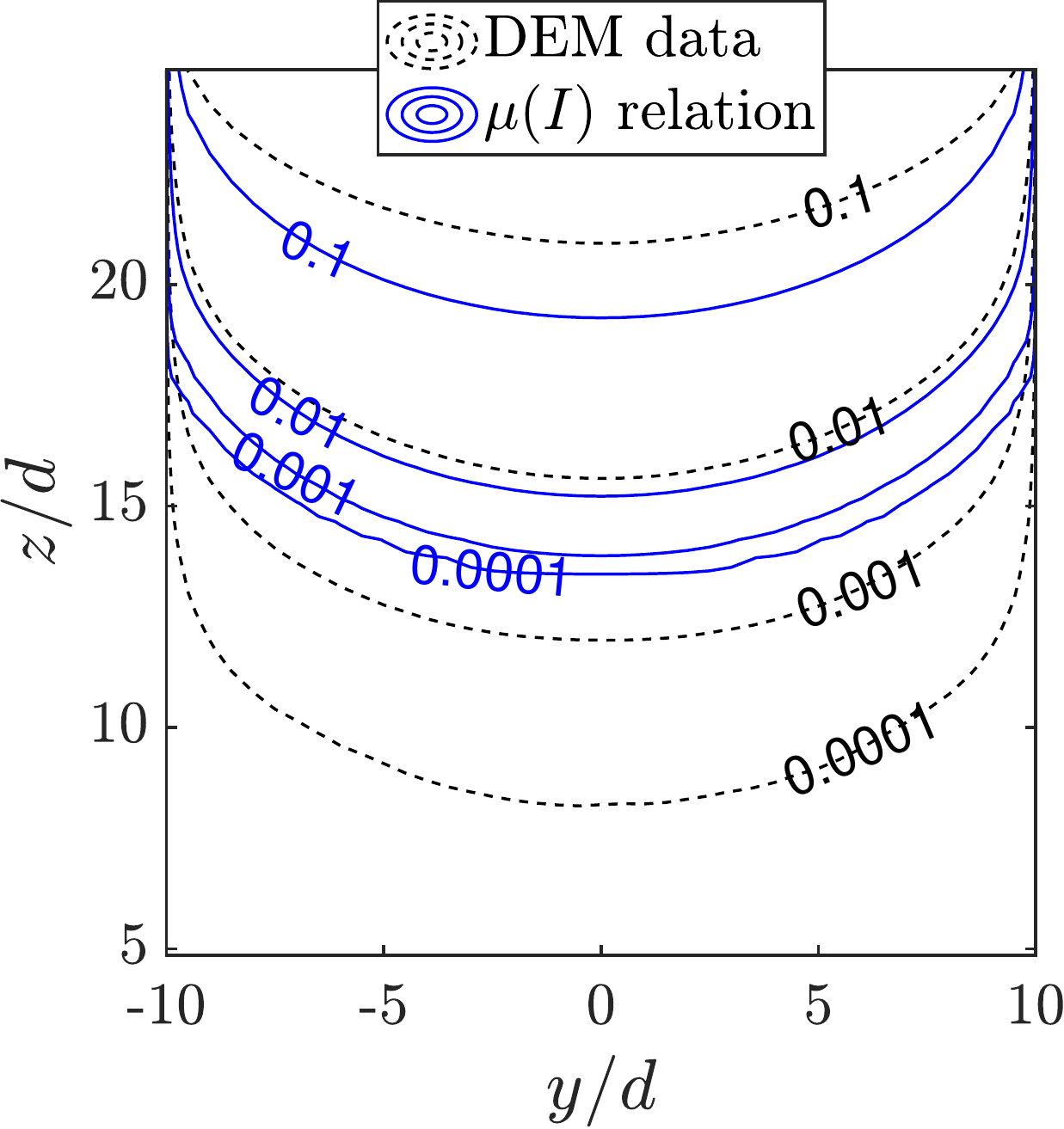}}
        \label{fig:3e}
    \end{subfigure}
    \quad
    \begin{subfigure}[t]{0.22\textwidth}  
        \centering 
        \phantomcaption
        \stackinset{l}{-0.02\textwidth}{b}{1.0\textwidth}
        {(\thesubfigure)}
        {\includegraphics[width=\textwidth]{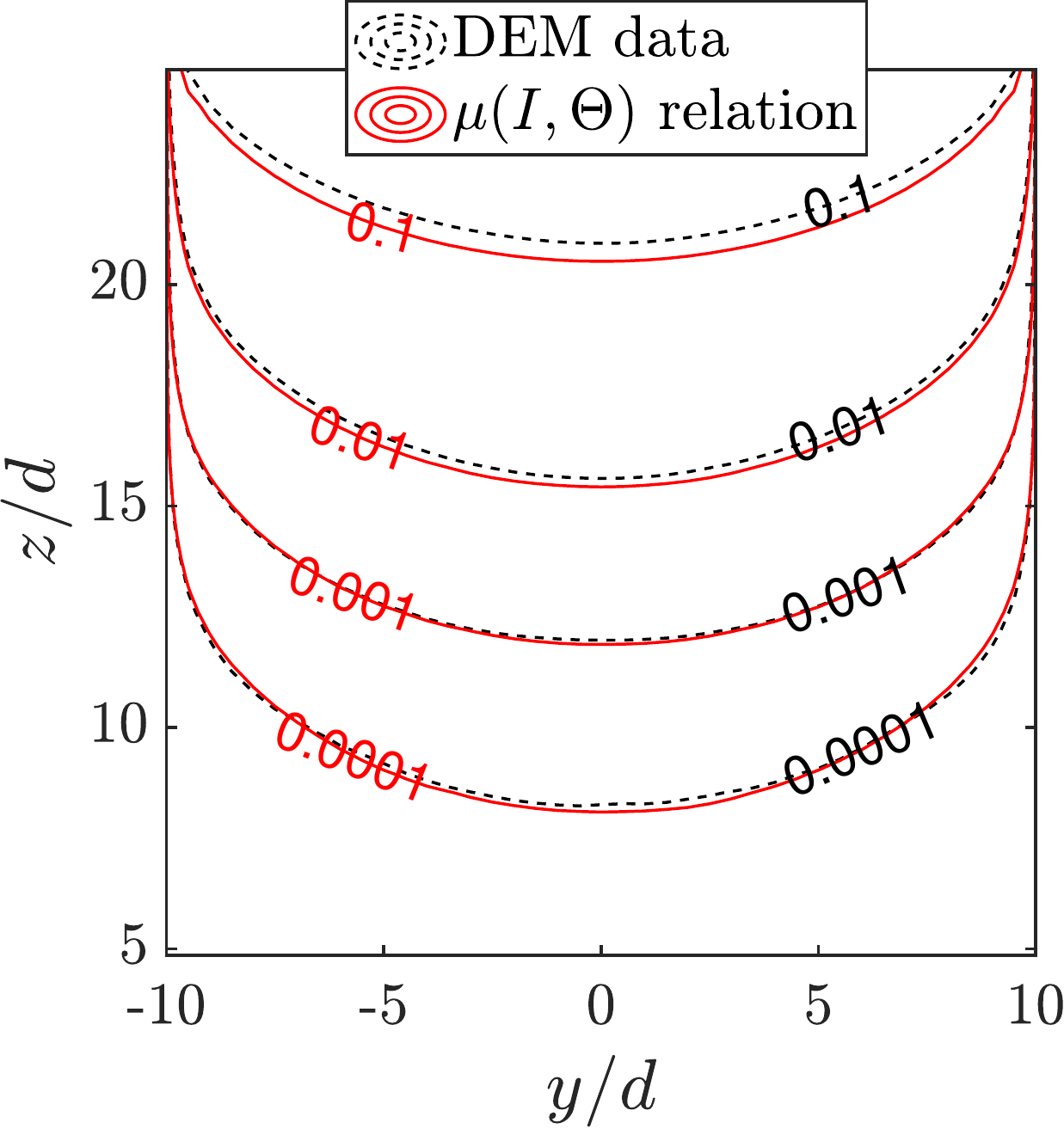}}
        \label{fig:3f}
    \end{subfigure}
    \captionsetup{justification=raggedright}
    \caption{(\protect\subref{fig:3a}) Inclined chute geometry with no-slip sides and a rough floor; DEM velocity pictured. Red particles fixed. (\protect\subref{fig:3b}) The distribution of $\log{\Theta}$.
    (\protect\subref{fig:3c}) Comparing the $\mu(I,\Theta)$ trend line previously obtained in planar shear tests to this geometry. (\protect\subref{fig:3d}) Comparing velocity from continuum model solutions to DEM data, viewed down the chute's center-plane (fixed $y$). (\protect\subref{fig:3e}) Comparing DEM velocity contours to that obtained from solving the $\mu(I)$ relation, and (\protect\subref{fig:3f}) the $\mu(I,\Theta)$ relation. The unit of velocity is $({|\vec{G}|H})^{1/2}$ where $H=20d$. All figures are for $\tan{\theta}=0.5$ except (\protect\subref{fig:3c}). See Supplemental Material for $\tan{\theta}=0.6$ case.
    }
    \label{fig:3}
\end{figure}

Next, we run simulations on inclined chute flows where the velocity depends on two spatial coordinates, $y$ and $z$ to check the predictive values of our $\mu(I,\Theta)$ rheology in a complex geometry. For easy calculations, we impose the no-slip boundary condition by setting two identical granular systems flowing in opposite directions periodically neighboring each other as in ~\cite{Chaudhuri2012} (see Supplemental Video).
We perform DEM simulations in a cuboid domain($L_x=120d$ and $L_y=40d$) using the same material used for the planar shear flows of 3D spheres with $\mu_p=0.4$ (see Fig.~\ref{fig:3a}). About $1.2\times10^5$ particles are simulated in total. The continuum fields are averaged along 300 lines (50 $y$ coordinates $\times$ 60 $z$ coordinates) parallel to the $x$ axis. The overlap lengths between the lines and the particles are used for the weighting in the coarse-graining.
We use a basis aligned with the local shearing planes, per~\cite{Depken2006}, so that $\mu$, $I$, and $\Theta$ are defined the same way as before. The same cut off standards are used.
Figure \ref{fig:3c} shows that $\mu\Theta^{1/6}=f(I)$ still holds in the complex geometry without refitting. All the data from two flows with different inclinations, $\theta=\tan^{-1}(0.5)$ and $\tan^{-1}(0.6)$, collapse to the master curve from Fig.~\ref{fig:2c}.

We also calculate chute flow velocity fields under the $\mu(I,\Theta)$ model and the $\mu(I)$ model using the steady-state Cauchy momentum equation $\partial_j\sigma_{ij}+\rho_s\phi {G}_i= {0}$. For the weight density term we fix $\phi=0.60$, inferred from the mean height and DEM floor pressure. We assume the stress deviator and the strain-rate tensor are co-directional. The boundary conditions are traction-free on the free surface and $\vec{v}=0$ on the other three boundaries.  Rather than assume a fluctuation energy balance relation to model the temperature field, we use $\Theta(y,z)$ extracted from the DEM data (see Fig.~\ref{fig:3b}). See Supplemental Material for simulation details. The steady-state velocity profile predicted by the $\mu(I,\Theta)$ relation is almost identical to the DEM data in Fig.~\ref{fig:3d} and Fig.~\ref{fig:3f}. However, the $\mu(I)$ rheology, which assumes vanishing shear rate where $\mu<\mu_s$, disagrees with the DEM data as shown in Fig.~\ref{fig:3d} and Fig.~\ref{fig:3e}.

\begin{figure}
    \centering
    \begin{subfigure}[b]{0.228\textwidth}
        \centering
        \phantomcaption
        \stackinset{l}{0.82\textwidth}{b}{0.87\textwidth}
        {(\thesubfigure)}
        {\includegraphics[width=\textwidth]{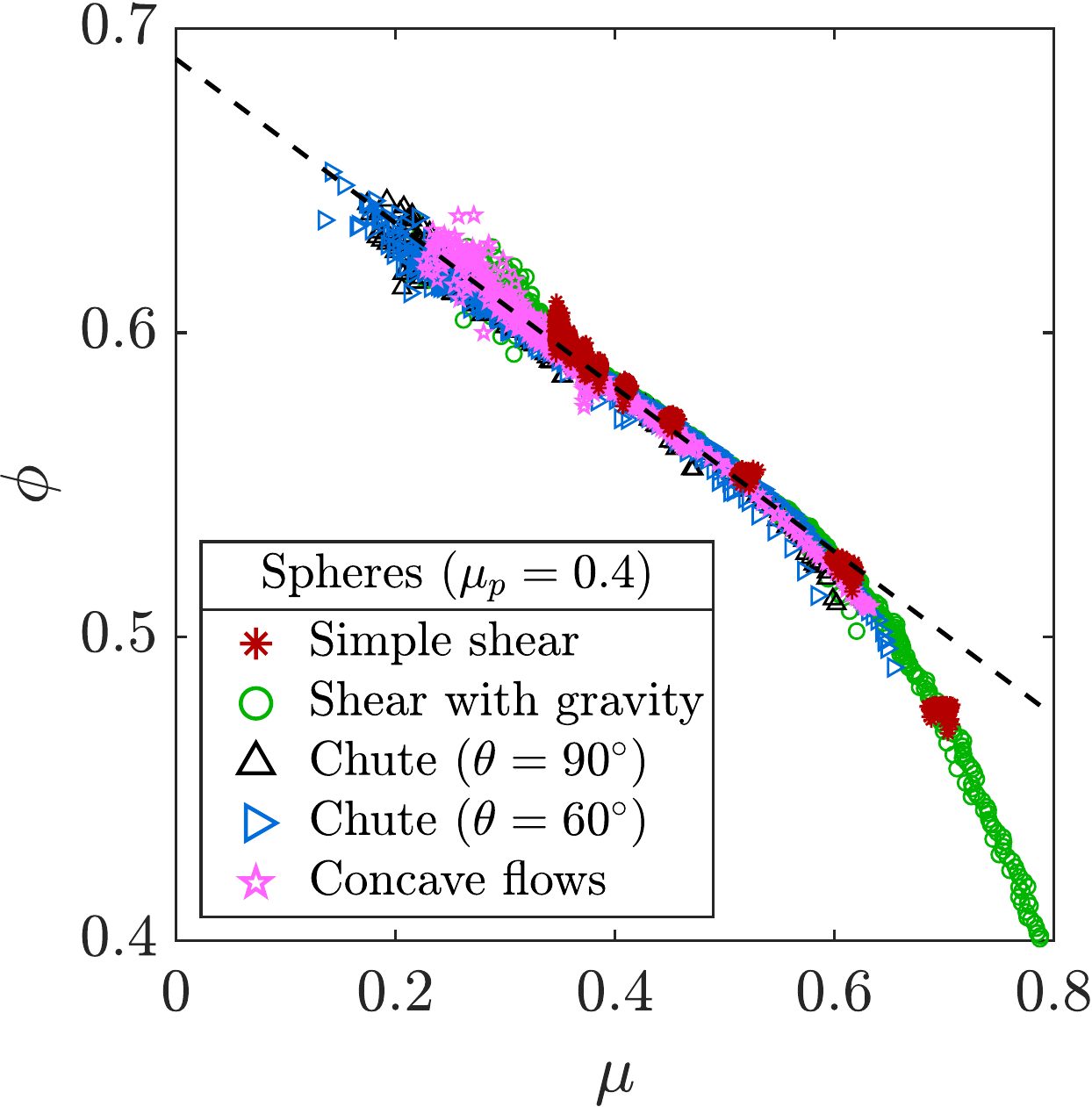}}
        \label{fig:4a}
    \end{subfigure}
    \quad
    \begin{subfigure}[b]{0.228\textwidth}  
        \centering 
        \phantomcaption
        \stackinset{l}{0.82\textwidth}{b}{0.87\textwidth}
        {(\thesubfigure)}
        {\includegraphics[width=\textwidth]{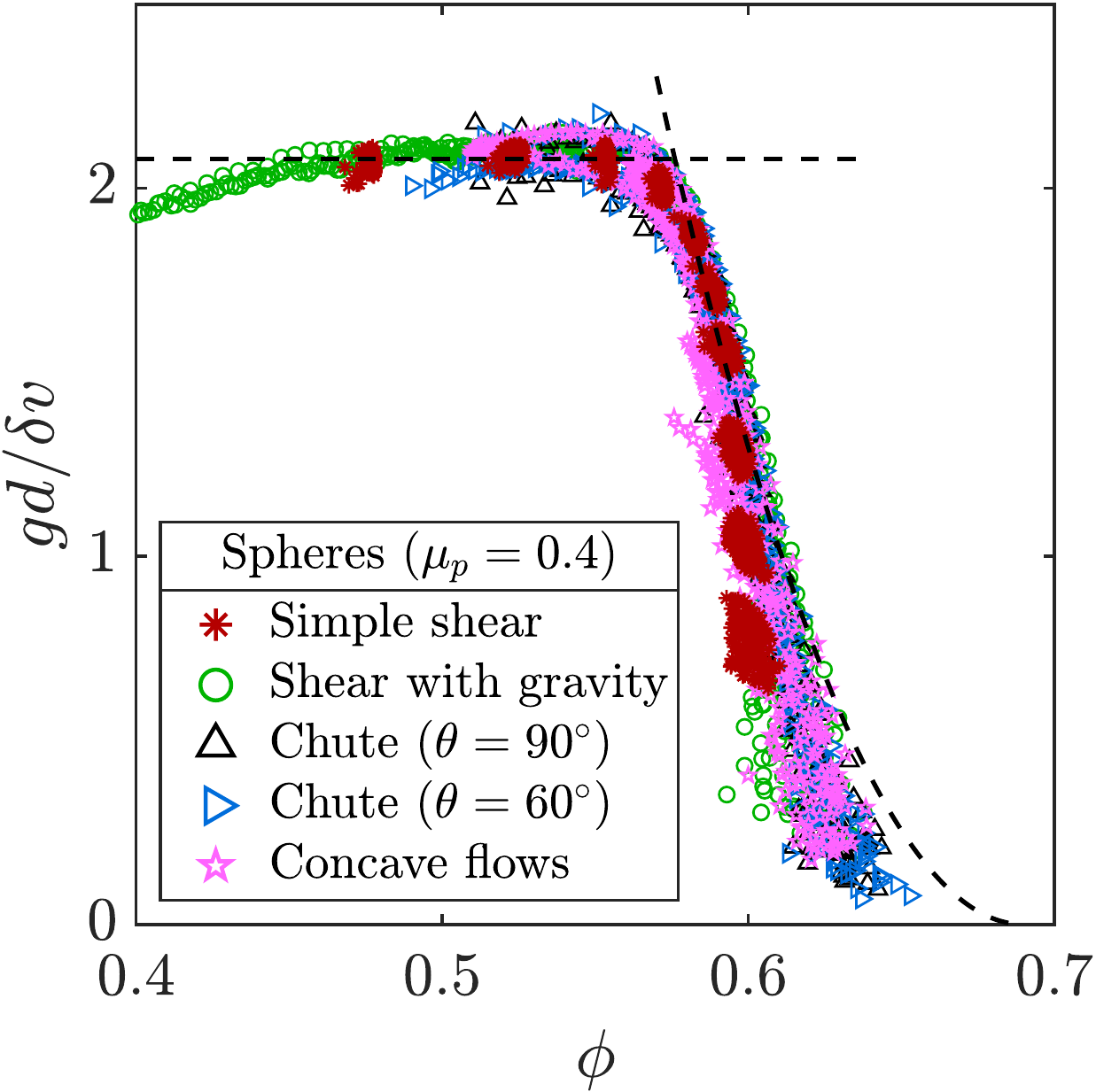}}
        \label{fig:4b}
    \end{subfigure}
    \captionsetup{justification=raggedright}
    \caption{DEM data in various planar shear flows of 3D spheres with $\mu_p=0.4$. (\protect\subref{fig:4a}) $\phi$ decreases linearly with $\mu$ for $\phi\gtrsim0.5$: $\phi\approx0.69-0.27\mu$ (dashed line).  (\protect\subref{fig:4b}) Our findings predict $\tilde{g}=gd/\delta v\sim(0.69-\phi)^2$ (dashed parabola) for $10^{-2.5}\lesssim I\lesssim 10^{-1}$ and $\tilde{g}\sim\text{constant}$ for $10^{-1}\lesssim I\lesssim 1$ (horizontal dashed line).}
    \label{fig:4}
\end{figure}

The connection between our rheology and the well-known $\mu(I)$ rheology becomes clearer when Eq.~\eqref{eq:1} is rewritten as
\begin{equation}\label{eq:2}
    \mu(I,\Theta)=\left(\frac{\Theta_{loc}(I)}{\Theta}\right)^p \mu_{loc}(I)
\end{equation}
where $\mu_{loc}(I)$ and $\Theta_{loc}(I)$ are $\mu$ and $\Theta$, respectively, locally determined by $I$ in simple shear flows. The $\mu(I)$ rheology is retrieved when $\Theta = \Theta_{loc}(I)$. Equation \eqref{eq:2} indicates the model can be calibrated entirely from simple shear tests, if $p$ is indeed universal and known for a family of materials. Additionally, Eq.~\eqref{eq:2} reflects the key physical idea that $\Theta$ produces fluidization; higher $\Theta$ scales down the flow strength at fixed $I$. The $\Theta$ field produces fluidization while presumably spreading diffusively due to an underlying fluctuation energy balance law governing the temperature ~\cite{Jenkins1983, Lun1984, Bocquet2001}; this bears a strong similarity with the dynamics/role of the NGF fluidity field, furthering the possibility of a connection between NGF's fluidity diffusion equation and fluctuation energy balance~\cite{Kamrin2019}.

Another consequential relation identified in our DEM simulations is a one-to-one relation between $\phi$ and $\mu$ (Fig.~\ref{fig:4a}) at steady state. Contrary to the standard kinetic theory where $\phi$ is determined by $\Theta$, it is not $\Theta$ but $\mu$ that collapses our $\phi$ data the best. In 3D developed flow, the packing fraction follows $\phi(\mu)\approx\phi_0-a\mu$ for $\phi\gtrsim0.5$ where $\phi_0=0.69$ and $a=0.27$. The same formula applies in 2D with $\phi_0=0.87$ and $a=0.22$ for $\phi\gtrsim0.78$. The effect of particle surface friction on the $\phi(\mu)$ relation is not large, confirming previous observations~\cite{daCruz2005}.

This $\phi(\mu)$ relation explains how our $\mu(I,\Theta)$ relation is connected to Zhang's fluidity expression $\tilde{g}\equiv g d/\delta v=F(\phi)$, which has been observed to hold in previous studies \cite{Zhang2017, Berzi2018, Qi2020}. First, we divide the range of $I$ into three regimes based on the slope of the master curve in Fig.~\ref{fig:2c}:
\begin{equation} \label{eq:3}
\mu\Theta^{1/6}=f(I)\sim
\begin{cases}
    I^{1/4} &\text{for $10^{-4}\lesssim I\lesssim 10^{-2.5}$}\\
    I^{1/3} &\text{for $10^{-2.5}\lesssim I\lesssim 10^{-1}$}\\
    I^{1/2} &\text{for $10^{-1}\lesssim I\lesssim 1$}.
\end{cases}
\end{equation}

 In the $1/2$ regime, $\Theta$ is mainly determined by $I$ following $\Theta \sim I^{3/2}$ (Fig.~\ref{fig:2b}). Combining this fact with 
Eq.~\eqref{eq:3} and the fact that $\tilde{g}$ can be rewritten as $I/\mu\sqrt{3\Theta}$, we obtain $\tilde{g}\sim \text{constant}$. This plateau regime is in line with kinetic theory where $\tilde{g}={F_1(\phi)}/\sqrt{3}{F_2(\phi)}$ becomes almost constant for $\phi\gtrsim 0.49$ ~\cite{Jenkins1983, Lun1984, Garzo1999, Jenkins2010, Berzi2018, Berzi2020}.
In the $1/3$ regime, $\Theta$ cancels out in the expression $\tilde{g}=I/\mu\sqrt{3\Theta}$ upon applying Eq.~\eqref{eq:3}, resulting in $\tilde{g}\sim\mu^2$, which can be further re-expressed under the linear $\phi(\mu)$ collapse as $\tilde{g}\sim(\phi_0-\phi)^2$.
Therefore, in the 1/3 regime, $\tilde{g}$ decreases quadratically in $\phi$.
Merging this regime's behavior with the plateau of the 1/2 regime, as shown in Fig.~\ref{fig:4}, delivers the basic large-$\tilde{g}$ behavior of the $\tilde{g}-\phi$ relationship apparent in our data and observed in \cite{Zhang2017}.
However, in the $1/4$ regime, corresponding to the lowest part ($\tilde{g}\lesssim1$) in Fig.~\ref{fig:4b}, it is clear from the data spread that Zhang's representation loses accuracy. The $\mu(I,\Theta)$ relation, on the other hand, remains well-collapsed and explains the spread in Zhang's representation as due to $\tilde{g}$ gaining additional $\Theta$ dependence; in the 1/4 regime, Eq.~\eqref{eq:3} and $\tilde{g}=I/\mu\sqrt{3\Theta}$ imply $\tilde{g}\sim (\phi_0-\phi)^3\Theta^{1/6}$.

Gaume and coworkers~\cite{Gaume2011} have also treated $\mu$, $I$, and $\Theta$ as independent variables to attempt a relation between them. They have suggested $\Theta \propto I^{h(\mu)}$ where $h(\mu)$ linearly changes with $\mu$. Although this formula approximately fits their DEM data in annular shear flow, our data does not match this trend and it appears their formula cannot be carried accurately to large $I$; $\mu$ is not determined at $I=1$, and $\mu$ increases as $\Theta$ increases for $I>1$.
 By comparison, advantages of our model include a form motivated by power-law scaling in phase transitions, covering up to higher $I$ and producing a strong data collapse over a wide array of geometries.  Our model also reveals a potentially universal scaling exponent $p$, which, once identified, allows model fitting solely from simple shear data using Eq.~\eqref{eq:2}. Additionally, our model offers a connection to and expansion from existing approaches, namely kinetic theory and the NGF model,  while clearly encapsulating, through Eq.~\eqref{eq:2}, the physical role of heat-softening.

Using many DEM simulations, we have found a general constitutive equation for simple granular materials, which relates three dimensionless variables: $\mu$, $I$, and $\Theta$. The granular rheology can be expressed as a power-law scaling form $\mu\Theta^p=f(I)$ where the exponent $p$ is about $1/6$ for 3D spheres and $1/8$ for 2D disks. $f(I)$ has certain general behaviors but details depend on the material properties. Our calibrated relation can be used to generate the velocity field in inclined chutes where flow depends on two spatial coordinates. We also observe a one-to-one relation between $\phi$ and $\mu$, which allows us to reconcile our model with $\phi$-dependent constitutive relations proposed by both the empirical and theoretical approaches. Kinetic theory, NGF modeling, and our current work all point strongly to the idea that the diffusing field responsible for granular nonlocality is directly related to the temperature. A clear next step is to explore the inclusion of a fluctuational energy balance law accurate into the dense regime; this would provide $\Theta$ and complete the rheological model.



%

\pagebreak
\renewcommand{\thepage}{S\arabic{page}} 
\renewcommand{\thesection}{S\arabic{section}}  
\renewcommand{\thetable}{S\arabic{table}}  
\renewcommand{\thefigure}{S\arabic{figure}}
\setcounter{page}{1}
\setcounter{section}{0}
\setcounter{table}{0}
\setcounter{figure}{0}

\widetext
\begin{center}
\vspace*{5mm}
\textbf{\large Supplemental Material for\\``Power-law scaling in granular rheology across flow geometries''}
\end{center}

\section{Simulation conditions}
We use LAMMPS, which implements the discrete element method (DEM), to simulate granular flows of 3D spheres and 2D disks. For the contact forces, we use the standard spring-dashpot model where the normal force is $F_{n} = k_n\delta_n - \gamma_n v_n$ and the tangential force is $F_{t} = k_t\delta_t$ where $\delta_n$ and $\delta_t$ are the normal and tangential components of the contact displacement respectively and ${v}_n$ is the normal component of the relative velocity. The tangential elastic constant $k_t$ is set to be $2/7$ times of the normal elastic constant $k_n$. The restitution coefficient $\epsilon$ is chosen to be $0.24$. The damping coefficient is then given by $\gamma_n = \sqrt{{2mk_n}/\left(1+(\pi/\ln{\epsilon})^2\right)}$~\cite{daCruz2005, Liu2018}. The simulation time step is set to be 6\% of the binary collision time $\tau_c=\pi\sqrt{\frac{m}{2k_n}\left(1+(\ln{\epsilon}/{\pi})^2\right)}$. The external body force in the concave flows is ${\vec F_z}=(mG/d)(z-z_0)\hat{z}$ where $G$ is a constant and $z_0$ is the midpoint of the system.

Table \ref{tab:1} to \ref{tab:4} summarize the simulation conditions. $N$ is the total number of particles except wall particles. The unit of pressure $P_0$ is $3.1\times10^{-7} k_n/d$ in 3D and $3.1\times10^{-7} k_n$ in 2D. The unit of acceleration $G_0$ is $(1/50)P_0/\rho_s$ in 3D and $(1/75)P_0/\rho_s d$ in 2D. The unit of velocity $V_0$ is $8.8\sqrt{P_0/\rho_s}$ in 3D and $4.8\sqrt{P_0/\rho_s}$ in 2D. We output data every $\Delta n$ steps to obtain total $N_{out}$ snapshots.
\\
\\

\setlength{\tabcolsep}{10pt} 
\renewcommand{\arraystretch}{1.0} 
\begin{table}[H]
    \centering
    \caption{Simulation conditions for the planar shear tests (3D spheres with $\mu_p=0.4$)}
    \begin{tabular}{ccccccc}
    \hline
    \textbf{Geometry} & \textbf{$N$} & \textbf{$P_{wall}/P_0$} & \textbf{$G/G_0$} & \textbf{$V_{wall}/V_0$} & \textbf{$\Delta n$} & \textbf{$N_{out}$} \\
    \hline
    Simple shear & 18327 & 4     &       & 0.003125 & 160000 & 1800 \\
    Simple shear & 18327 & 4     &       & 0.0125 & 80000 & 1800 \\
    Simple shear & 18327 & 4     &       & 0.05  & 40000 & 1800 \\
    Simple shear & 18327 & 1     &       & 0.1   & 40000 & 1800 \\
    Simple shear & 18327 & 1     &       & 0.2   & 40000 & 1800 \\
    Simple shear & 18327 & 1     &       & 0.4   & 40000 & 1800 \\
    Simple shear & 18327 & 1     &       & 0.8   & 40000 & 1800 \\
    Simple shear & 18327 & 1     &       & 1.6   & 40000 & 1800 \\
    Simple shear & 18327 & 1     &       & 3.2   & 20000 & 3600 \\
    Simple shear & 18327 & 1     &       & 6.4   & 20000 & 3600 \\
    Shear with gravity & 18327 & 1     & 16    & 3.2   & 20000 & 7200 \\
    Shear with gravity & 18327 & 1     & 2     & 12.8  & 20000 & 7200 \\
    Shear with gravity & 18327 & 1     & 32    & 1.6   & 40000 & 3600 \\
    Shear with gravity & 18327 & 1     & 4     & 6.4   & 40000 & 7200 \\
    Shear with gravity & 18327 & 1     & 8     & 1.6   & 40000 & 3600 \\
    Shear with gravity & 18327 & 4     & 8     & 0.1   & 40000 & 3600 \\
    Chute flows ($\theta=90^\circ$) & 18327 & 8     & 12    &       & 40000 & 3600 \\
    Chute flows ($\theta=90^\circ$) & 18327 & 8     & 16    &       & 40000 & 3600 \\
    Chute flows ($\theta=90^\circ$) & 18327 & 8     & 20    &       & 40000 & 3600 \\
    Chute flows ($\theta=60^\circ$) & 18327 & 8     & 16    &       & 40000 & 3600 \\
    Chute flows ($\theta=60^\circ$) & 18327 & 8     & 20    &       & 40000 & 3600 \\
    Chute flows ($\theta=60^\circ$) & 18327 & 8     & 24    &       & 40000 & 3600 \\
    Concave flows & 18327 & 16    & 3.5   & 0.00625 & 80000 & 3600 \\
    Concave flows & 18327 & 16    & 3     & 0.025 & 40000 & 3600 \\
    Concave flows & 18327 & 16    & 3     & 0.4   & 40000 & 3600 \\
    Concave flows & 18327 & 16    & 3     & 1.6   & 20000 & 7200 \\
    \hline
    \end{tabular}%
    \label{tab:1}%
\end{table}%

\begin{table}[H]
  \centering
    \caption{Simulation conditions for the planar shear tests (3D spheres with $\mu_p=0.1$)}
    \begin{tabular}{ccccccc}
 \hline
    \textbf{Geometry} & \textbf{$N$} & \textbf{$P_{wall}/P_0$} & \textbf{$G/G_0$} & \textbf{$V_{wall}/V_0$} & \textbf{$\Delta n$} & \textbf{$N_{out}$} \\
    \hline
    Simple shear & 18327 & 4     &       & 0.0015625 & 160000 & 1800 \\
    Simple shear & 6923  & 4     &       & 0.0015625 & 160000 & 3600 \\
    Simple shear & 6923  & 4     &       & 0.003125 & 160000 & 1800 \\
    Simple shear & 6923  & 4     &       & 0.0125 & 80000 & 1800 \\
    Simple shear & 6923  & 4     &       & 0.05  & 40000 & 1800 \\
    Simple shear & 6923  & 1     &       & 0.1   & 40000 & 1800 \\
    Simple shear & 6923  & 1     &       & 0.2   & 40000 & 1800 \\
    Simple shear & 6923  & 1     &       & 0.4   & 40000 & 1800 \\
    Simple shear & 6923  & 1     &       & 0.8   & 40000 & 1800 \\
    Simple shear & 6923  & 1     &       & 1.6   & 40000 & 1800 \\
    Simple shear & 6923  & 1     &       & 3.2   & 40000 & 3600 \\
    Simple shear & 6923  & 1     &       & 6.4   & 40000 & 3600 \\
    Shear with gravity & 6923  & 1     & 1     & 6.4   & 20000 & 14400 \\
    Shear with gravity & 18327 & 1     & 16    & 0.1   & 40000 & 3600 \\
    Shear with gravity & 18327 & 1     & 16    & 1.6   & 40000 & 3600 \\
    Shear with gravity & 18327 & 1     & 4     & 1.6   & 40000 & 3600 \\
    Chute flows ($\theta=90^\circ$) & 18327 & 8     & 10    &       & 40000 & 3600 \\
    Chute flows ($\theta=90^\circ$) & 18327 & 8     & 12    &       & 40000 & 3600 \\
    Chute flows ($\theta=90^\circ$) & 18327 & 8     & 14    &       & 40000 & 3600 \\
    Chute flows ($\theta=60^\circ$) & 18327 & 8     & 10    &       & 40000 & 3600 \\
    Chute flows ($\theta=60^\circ$) & 18327 & 8     & 12    &       & 40000 & 3600 \\
    Chute flows ($\theta=60^\circ$) & 18327 & 8     & 14    &       & 40000 & 3600 \\
    Concave flows & 18327 & 16    & 3.5   & 0.025 & 40000 & 3600 \\
    Concave flows & 18327 & 16    & 3     & 0.4   & 40000 & 3600 \\
    Concave flows & 18327 & 16    & 3     & 1.6   & 20000 & 7200 \\
    \hline
    \end{tabular}%
  \label{tab:2}%
\end{table}%

\begin{table}[H]
  \centering
    \caption{Simulation conditions for the planar shear tests (2D disks with $\mu_p=0.4$)}
    \begin{tabular}{ccccccc}
 \hline
    \textbf{Geometry} & \textbf{$N$} & \textbf{$P_{wall}/P_0$} & \textbf{$G/G_0$} & \textbf{$V_{wall}/V_0$} & \textbf{$\Delta n$} & \textbf{$N_{out}$} \\
    \hline
    Simple shear & 6739  & 1     &       & 0.0015625 & 160000 & 3600 \\
    Simple shear & 6739  & 1     &       & 0.00625 & 80000 & 3600 \\
    Simple shear & 6739  & 1     &       & 0.025 & 20000 & 3600 \\
    Simple shear & 6739  & 1     &       & 0.1   & 20000 & 3600 \\
    Simple shear & 6739  & 1     &       & 0.2   & 20000 & 3600 \\
    Simple shear & 6739  & 1     &       & 0.4   & 20000 & 3600 \\
    Simple shear & 6739  & 1     &       & 0.8   & 20000 & 3600 \\
    Simple shear & 6739  & 1     &       & 1.6   & 20000 & 3600 \\
    Simple shear & 6739  & 1     &       & 3.2   & 10000 & 7200 \\
    Simple shear & 6739  & 1     &       & 6.4   & 10000 & 7200 \\
    Simple shear & 6739  & 1     &       & 12.8  & 40000 & 7200 \\
    Simple shear & 6739  & 1     &       & 25.6  & 40000 & 7200 \\
    Shear with gravity & 6739  & 1     & 1     & 12.8  & 20000 & 14400 \\
    Shear with gravity & 6739  & 1     & 1     & 6.4   & 20000 & 14400 \\
    Shear with gravity & 19810 & 4     & 16    & 4.8   & 40000 & 3600 \\
    Shear with gravity & 19810 & 4     & 2     & 0.075 & 40000 & 3600 \\
    Shear with gravity & 19810 & 4     & 2     & 4.8   & 40000 & 3600 \\
    Shear with gravity & 19810 & 4     & 4     & 0.3   & 40000 & 3600 \\
    Chute flows ($\theta=90^\circ$) & 19810 & 4     & 2     &       & 40000 & 3600 \\
    Chute flows ($\theta=90^\circ$) & 19810 & 4     & 3     &       & 40000 & 3600 \\
    Chute flows ($\theta=90^\circ$) & 19810 & 4     & 4     &       & 40000 & 3600 \\
    Chute flows ($\theta=60^\circ$) & 19810 & 4     & 3     &       & 40000 & 3600 \\
    Chute flows ($\theta=60^\circ$) & 19810 & 4     & 4     &       & 40000 & 3600 \\
    Chute flows ($\theta=60^\circ$) & 19810 & 4     & 6     &       & 40000 & 3600 \\
    Concave flows & 19810 & 16    & 2/3     & 0.15  & 40000 & 3600 \\
    Concave flows & 19810 & 16    & 2/3     & 1.2   & 40000 & 3600 \\
    Concave flows & 19810 & 16    & 2/3     & 4.8   & 40000 & 3600 \\
    Concave flows & 19810 & 4     & 1/6   & 2.4   & 40000 & 3600 \\
    \hline
    \end{tabular}%
  \label{tab:3}%
\end{table}%

\begin{table}[H]
  \centering
    \caption{Simulation conditions for the planar shear tests (2D disks with $\mu_p=0.1$)}
    \begin{tabular}{ccccccc}
    \hline
    \textbf{Geometry} & \textbf{$N$} & \textbf{$P_{wall}/P_0$} & \textbf{$G/G_0$} & \textbf{$V_{wall}/V_0$} & \textbf{$\Delta n$} & \textbf{$N_{out}$} \\
    \hline
    Simple shear & 6739  & 1     &       & 0.0015625 & 160000 & 3600 \\
    Simple shear & 6739  & 1     &       & 0.00625 & 80000 & 3600 \\
    Simple shear & 6739  & 1     &       & 0.025 & 40000 & 3600 \\
    Simple shear & 6739  & 1     &       & 0.1   & 20000 & 3600 \\
    Simple shear & 6739  & 1     &       & 0.2   & 20000 & 3600 \\
    Simple shear & 6739  & 1     &       & 0.4   & 20000 & 3600 \\
    Simple shear & 6739  & 1     &       & 0.8   & 20000 & 3600 \\
    Simple shear & 6739  & 1     &       & 1.6   & 20000 & 3600 \\
    Simple shear & 6739  & 1     &       & 3.2   & 20000 & 3600 \\
    Simple shear & 6739  & 1     &       & 6.4   & 20000 & 3600 \\
    Simple shear & 6739  & 1     &       & 12.8  & 20000 & 3600 \\
    Shear with gravity & 6739  & 1     & 16    & 0.1   & 40000 & 3600 \\
    Shear with gravity & 6739  & 1     & 1     & 12.8  & 20000 & 14400 \\
    Shear with gravity & 6739  & 1     & 1     & 6.4   & 20000 & 7200 \\
    Shear with gravity & 6739  & 1     & 8     & 0.4   & 40000 & 3600 \\
    Shear with gravity & 6739  & 4     & 4     & 0.025 & 40000 & 3600 \\
    Shear with gravity & 6739  & 4     & 4     & 0.1   & 40000 & 3600 \\
    Chute flows ($\theta=90^\circ$) & 19810 & 4     & 2     &       & 40000 & 3600 \\
    Chute flows ($\theta=90^\circ$) & 19810 & 4     & 3     &       & 40000 & 3600 \\
    Chute flows ($\theta=90^\circ$) & 19810 & 4     & 4     &       & 40000 & 3600 \\
    Chute flows ($\theta=60^\circ$) & 19810 & 4     & 2     &       & 40000 & 3600 \\
    Chute flows ($\theta=60^\circ$) & 19810 & 4     & 3     &       & 40000 & 3600 \\
    Chute flows ($\theta=60^\circ$) & 19810 & 4     & 4     &       & 40000 & 3600 \\
    Concave flows & 19810 & 16    & 2/3     & 0.075 & 40000 & 3600 \\
    Concave flows & 19810 & 16    & 2/3     & 0.3   & 40000 & 3600 \\
    Concave flows & 19810 & 16    & 2/3     & 4.8   & 40000 & 3600 \\
    \hline
    \end{tabular}%
  \label{tab:4}%
\end{table}%
\begin{table}[H]
  \centering
    \caption{Simulation conditions for the inclined chute flows (3D spheres with $\mu_p=0.4$)}
    \begin{tabular}{ccccccc}
    \hline
    \textbf{Geometry} & \textbf{$N$} & \textbf{$G/G_0$} & \textbf{$\tan{\theta}$} & \textbf{$\Delta n$} & \textbf{$N_{out}$} \\
    \hline
    Inclined chute flows & 115619 & 64     & 0.5   & 40000 & 3600 \\
    Inclined chute flows & 115619 & 64     & 0.6   & 40000 & 3600 \\
    \hline
    \end{tabular}%
  \label{tab:5}%
\end{table}%

\section{Fitting functions in Fig. 2}
We approximate the master curves as
\begin{equation*} \label{eq:S2}
f(I)\approx
\begin{cases}
    0.20I^{0.24} + 0.32I^{0.72} &\text{for 3D spheres with $\mu_p = 0.4$}\\
    0.14I^{0.24} + 0.30I^{0.68} &\text{for 3D spheres with $\mu_p = 0.1$}\\
    0.21I^{0.17} + 0.55I^{0.75} &\text{for 2D disks with $\mu_p = 0.4$}\\
    0.17I^{0.18} + 0.47I^{0.75} &\text{for 2D disks with $\mu_p = 0.1$}
\end{cases}
\end{equation*}
which are drawn in Fig. 2. The dashed line in Fig. 2a is $\mu_{loc}(I) \approx 0.35+0.36I^{0.88-0.42I^{0.25}}$, and the one in Fig. 2b is determined by $\Theta_{loc}(I) = (f(I)/\mu_{loc}(I))^6$.

\section{Supplemental Figures}
\begin{figure}[H]
    \centering
    \begin{subfigure}[t]{0.227\textwidth}
        \centering
        \phantomcaption
        \stackinset{l}{0.85\textwidth}{b}{0.17\textwidth}
        {(\thesubfigure)}
        {\includegraphics[width=\textwidth]{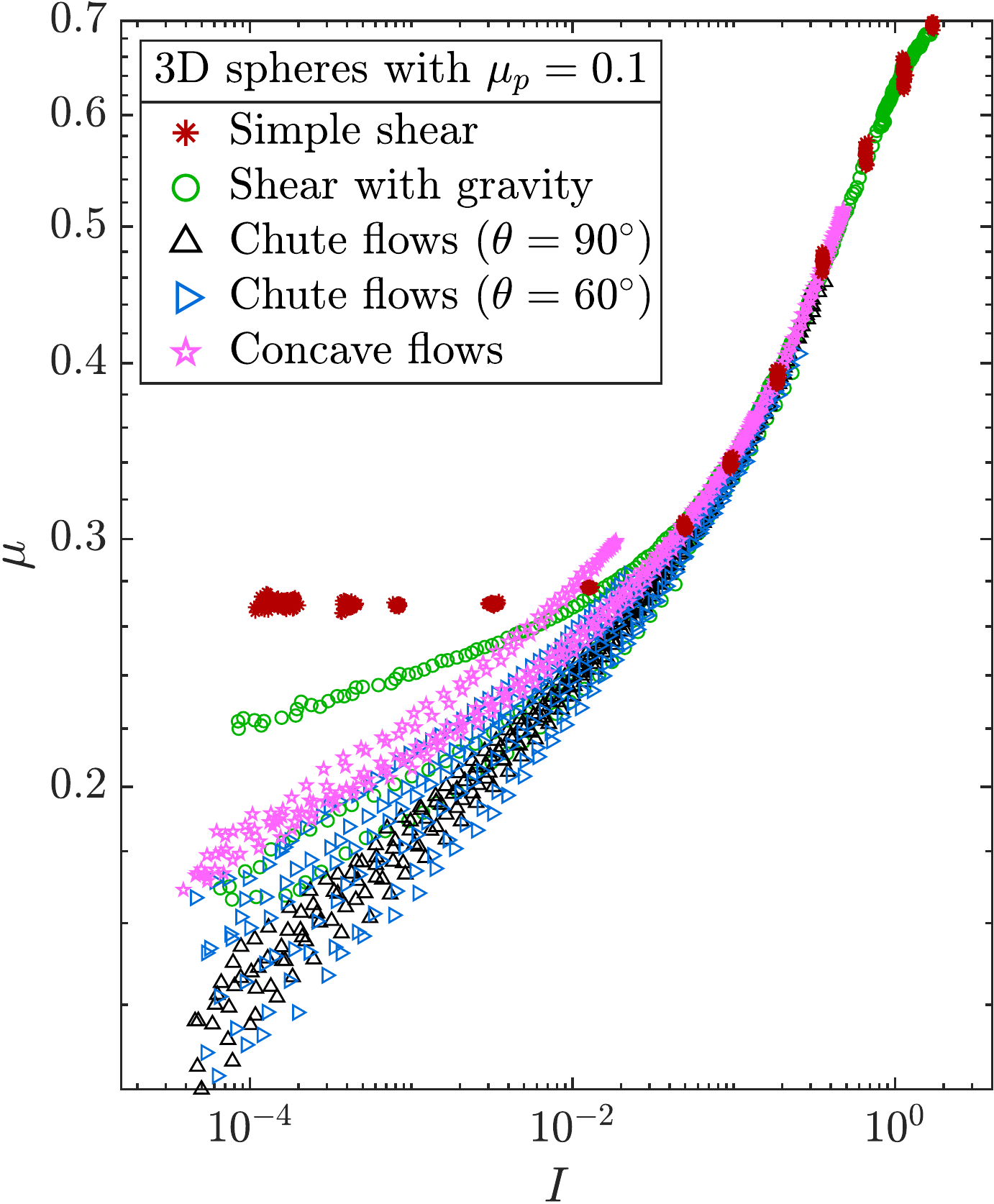}}
        \label{fig:S1a}
    \end{subfigure}
    \quad
    \begin{subfigure}[t]{0.232\textwidth}  
        \centering
        \phantomcaption
        \stackinset{l}{0.85\textwidth}{b}{0.17\textwidth}
        {(\thesubfigure)}
        {\includegraphics[width=\textwidth]{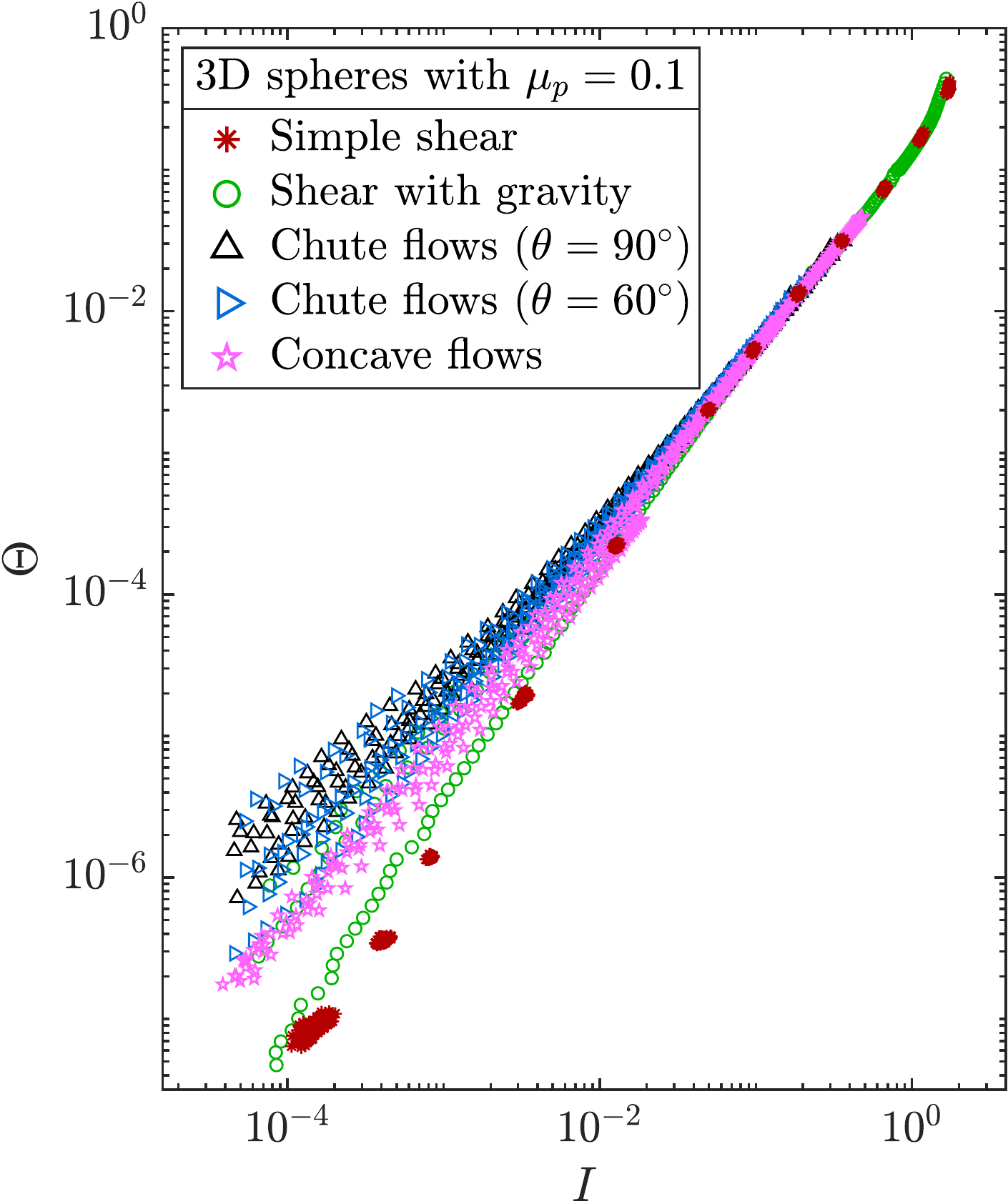}}
        \label{fig:S1b}
    \end{subfigure}
    \quad
    \begin{subfigure}[t]{0.227\textwidth}  
        \centering 
        \phantomcaption
        \stackinset{l}{0.85\textwidth}{b}{0.17\textwidth}
        {(\thesubfigure)}
        {\includegraphics[width=\textwidth]{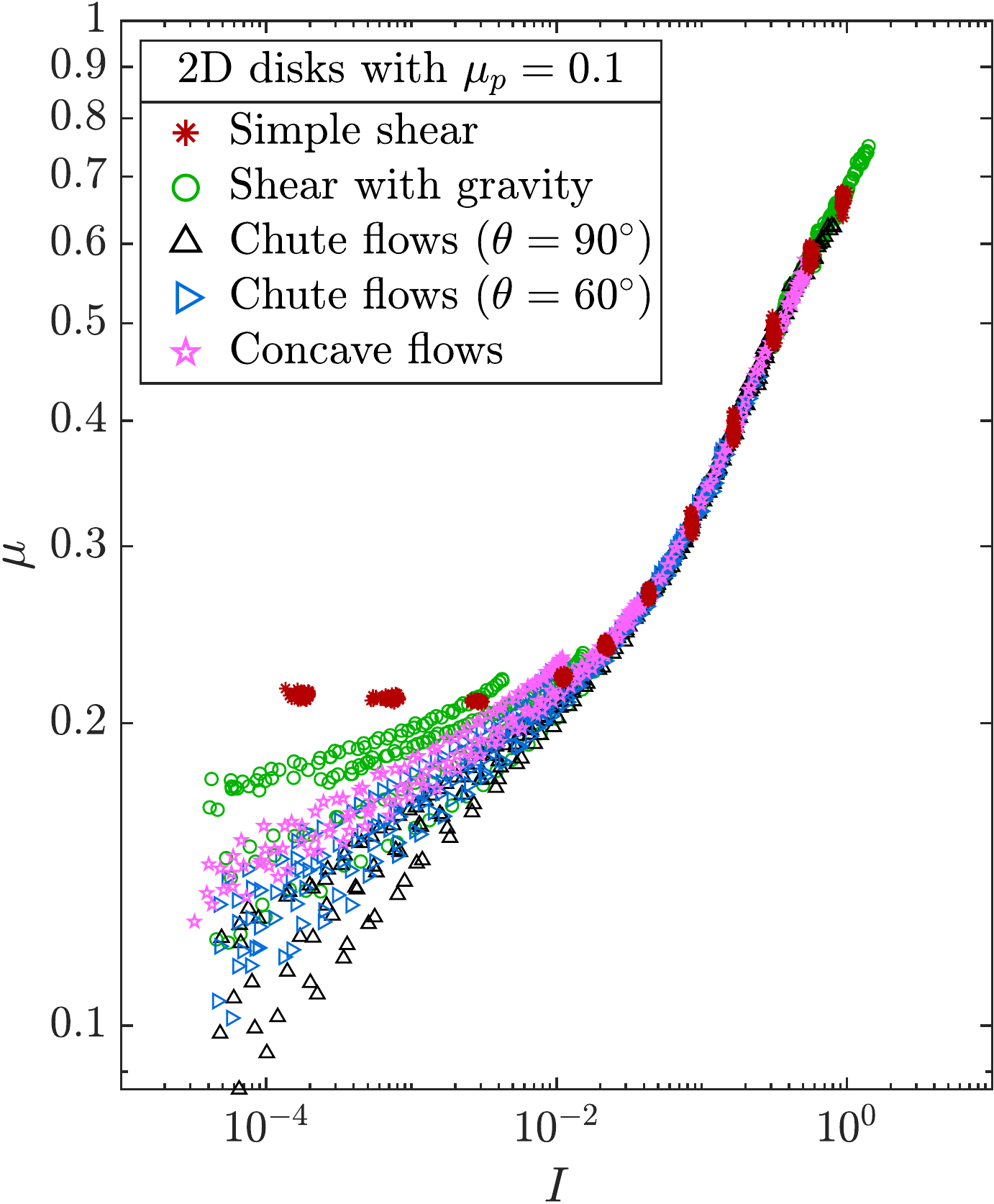}}
        \label{fig:S1c}
    \end{subfigure}
    \quad
    \begin{subfigure}[t]{0.232\textwidth}  
        \centering 
        \phantomcaption
        \stackinset{l}{0.85\textwidth}{b}{0.17\textwidth}
        {(\thesubfigure)}
        {\includegraphics[width=\textwidth]{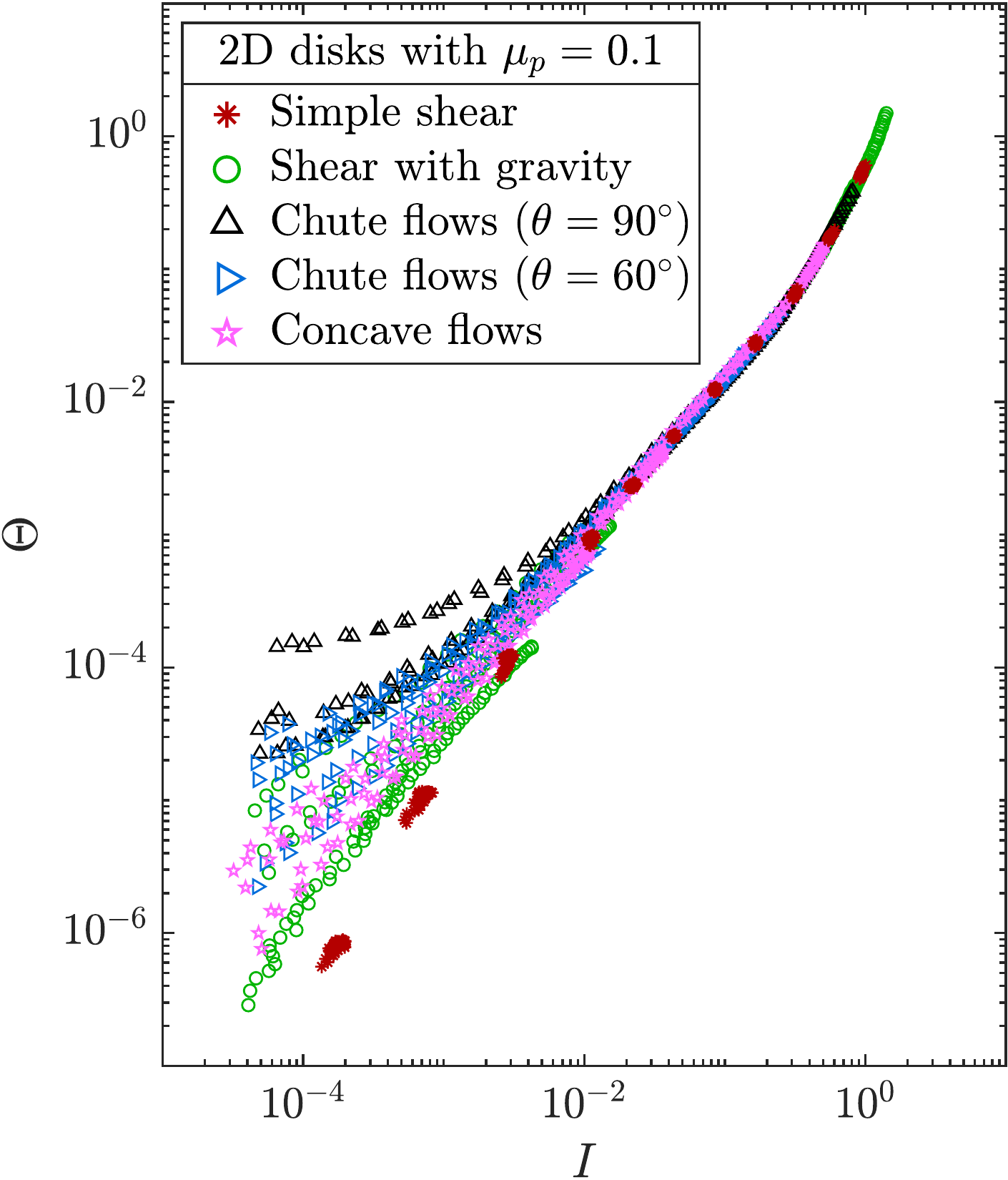}}
        \label{fig:S1d}
    \end{subfigure}
     \captionsetup{justification=raggedright, singlelinecheck=false}
    \caption{ DEM data from planar shear tests.
    Non-collapse of $\mu$ vs $I$ (\protect\subref{fig:S1a}) and $\Theta$ vs $I$ (\protect\subref{fig:S1b}) for 3D spheres with $\mu_p=0.1$.
    Similar non-collapse of $\mu$ vs $I$ (\protect\subref{fig:S1c}) and $\Theta$ vs $I$ (\protect\subref{fig:S1d}) for 2D disks with $\mu_p=0.1$.}
    \label{fig:S1}
\end{figure}

\begin{figure}[h]
    \centering
    \begin{subfigure}[t]{0.24\textwidth}
        \centering
        \phantomcaption
        \stackinset{l}{0.84\textwidth}{b}{1.03\textwidth}
        {(\thesubfigure)}
        {\includegraphics[width=\textwidth]{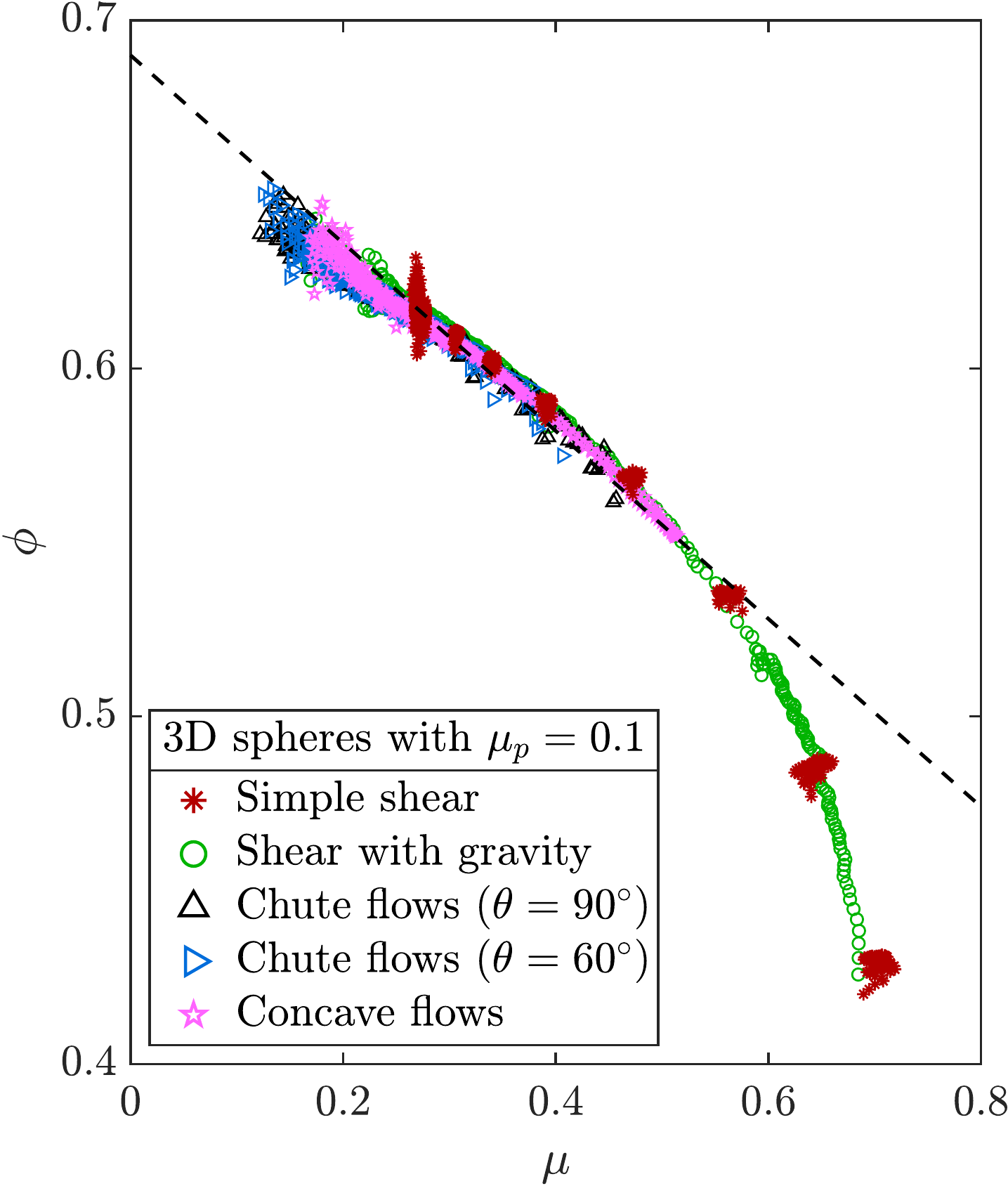}}
        \label{fig:S2a}
    \end{subfigure}
    \quad
    \begin{subfigure}[t]{0.24\textwidth}  
        \centering
        \phantomcaption
        \stackinset{l}{0.84\textwidth}{b}{1.03\textwidth}
        {(\thesubfigure)}
        {\includegraphics[width=\textwidth]{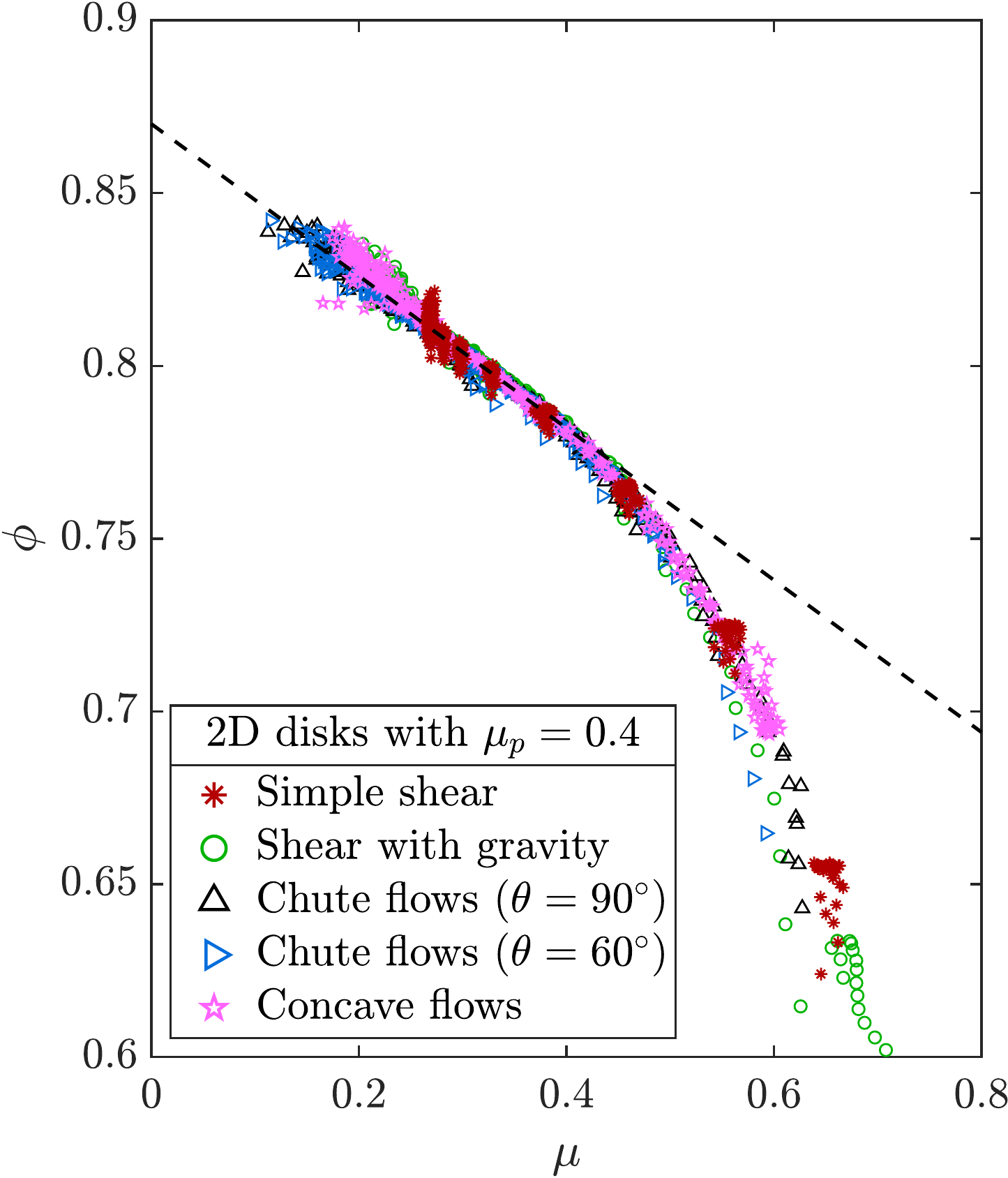}}
        \label{fig:S2b}
    \end{subfigure}
    \quad
    \begin{subfigure}[t]{0.24\textwidth}  
        \centering 
        \phantomcaption
        \stackinset{l}{0.84\textwidth}{b}{1.03\textwidth}
        {(\thesubfigure)}
        {\includegraphics[width=\textwidth]{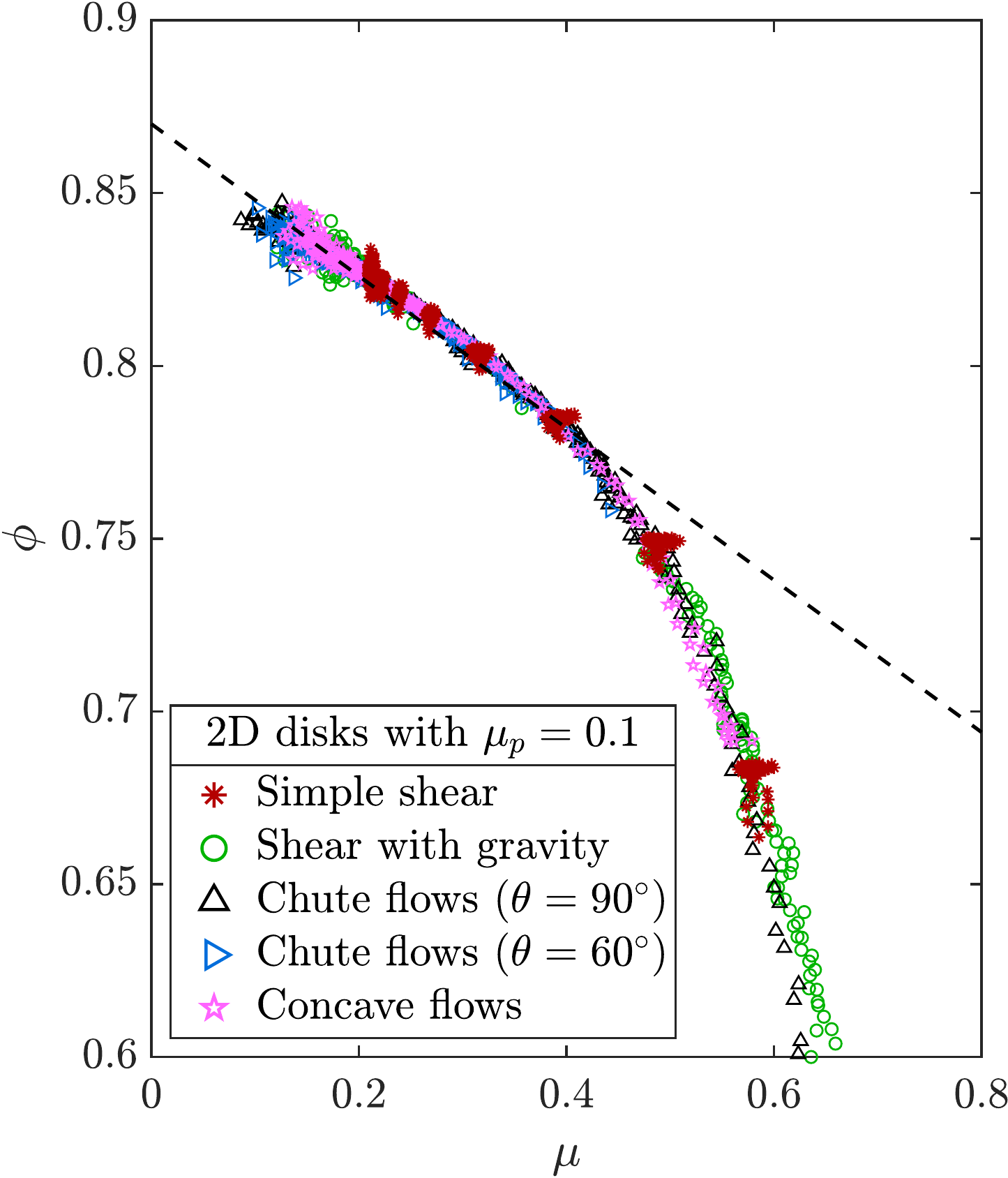}}
        \label{fig:S2c}
    \end{subfigure}
     \captionsetup{justification=raggedright, singlelinecheck=false}
    \caption{ DEM data from planar shear tests.
    One-to-one relationship between $\phi$ and $\mu$ for 3D spheres with $\mu_p=0.1$ (\protect\subref{fig:S2a}), 2D disks with $\mu_p=0.4$ (\protect\subref{fig:S2b}), and 2D disks with $\mu_p=0.1$ (\protect\subref{fig:S2c}).
    Dashed line in (\protect\subref{fig:S2a}) is the same trend line as Fig. 4a: $\phi=0.69-0.27\mu$.
    Dashed lines in (\protect\subref{fig:S2b}) and (\protect\subref{fig:S2c}) are the same: $\phi=0.87-0.22\mu$.}
    \label{fig:S2}
\end{figure}

\begin{figure}[H]
    \centering
    \begin{subfigure}[t]{0.228\textwidth}
        \centering
        \phantomcaption
        \stackinset{l}{0.84\textwidth}{b}{1.03\textwidth}
        {(\thesubfigure)}
        {\includegraphics[width=\textwidth]{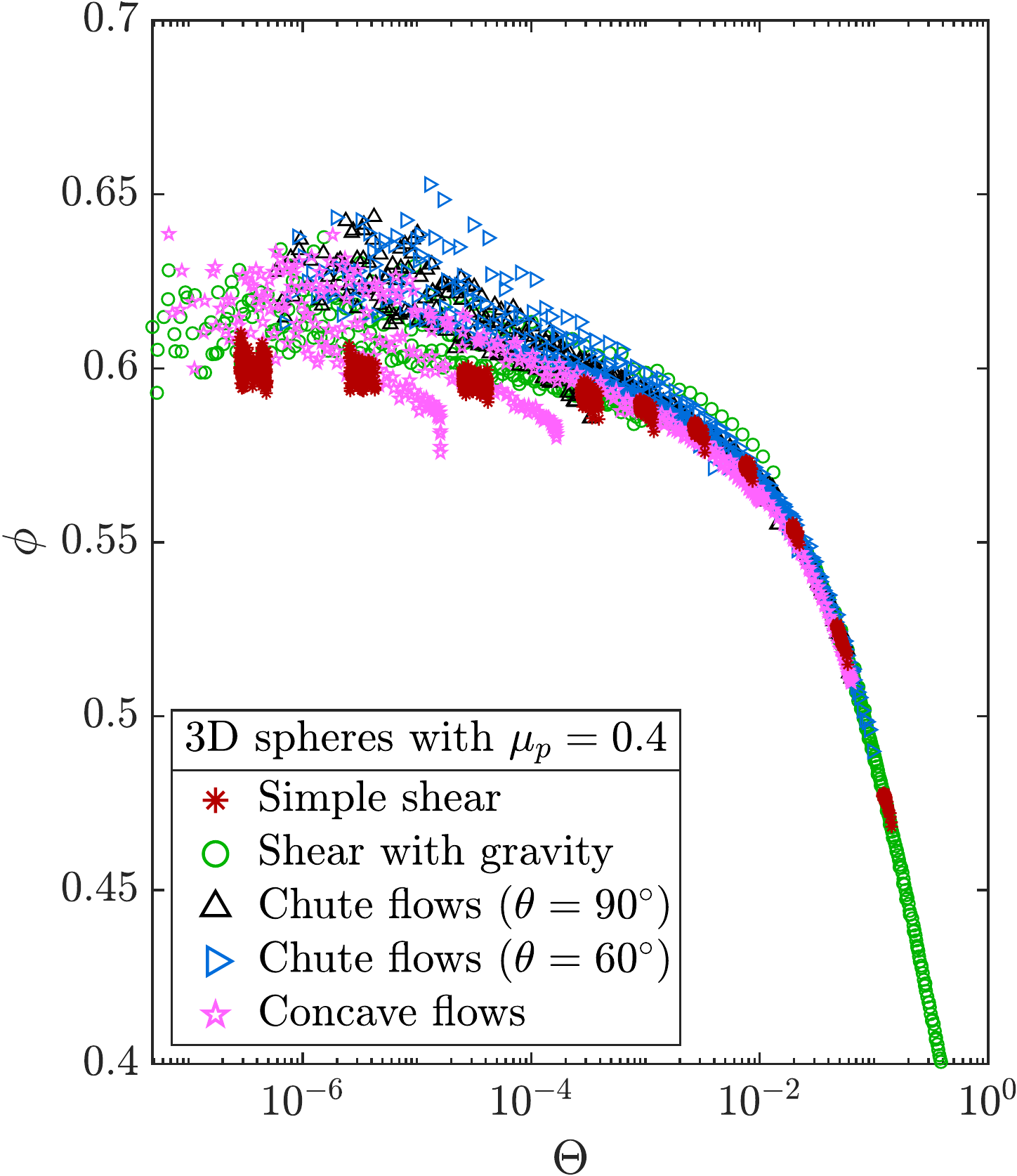}}
        \label{fig:S3a}
    \end{subfigure}
    \quad
    \begin{subfigure}[t]{0.228\textwidth}  
        \centering
        \phantomcaption
        \stackinset{l}{0.84\textwidth}{b}{1.03\textwidth}
        {(\thesubfigure)}
        {\includegraphics[width=\textwidth]{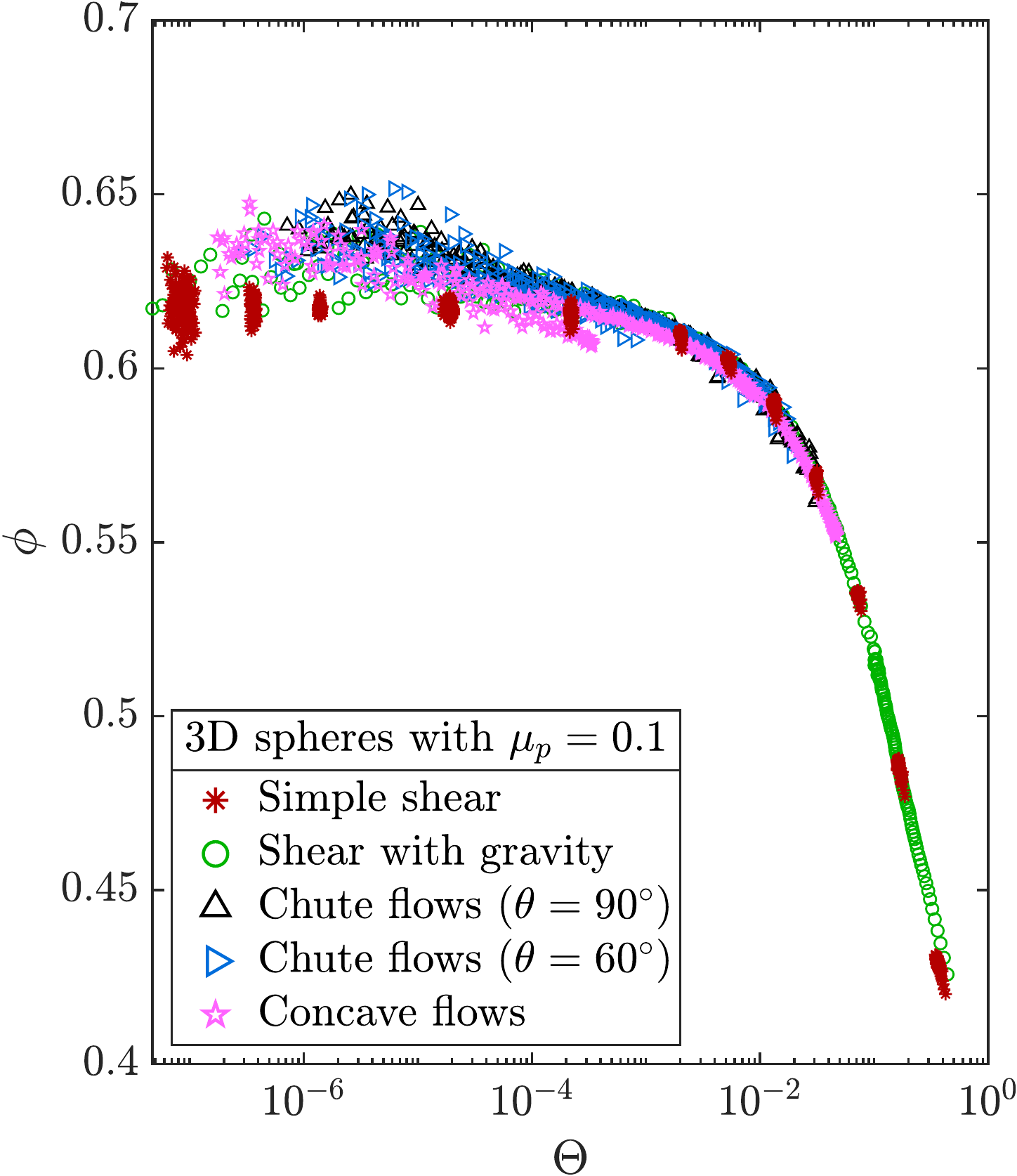}}
        \label{fig:S3b}
    \end{subfigure}
    \quad
    \begin{subfigure}[t]{0.232\textwidth}  
        \centering 
        \phantomcaption
        \stackinset{l}{0.84\textwidth}{b}{1.01\textwidth}
        {(\thesubfigure)}
        {\includegraphics[width=\textwidth]{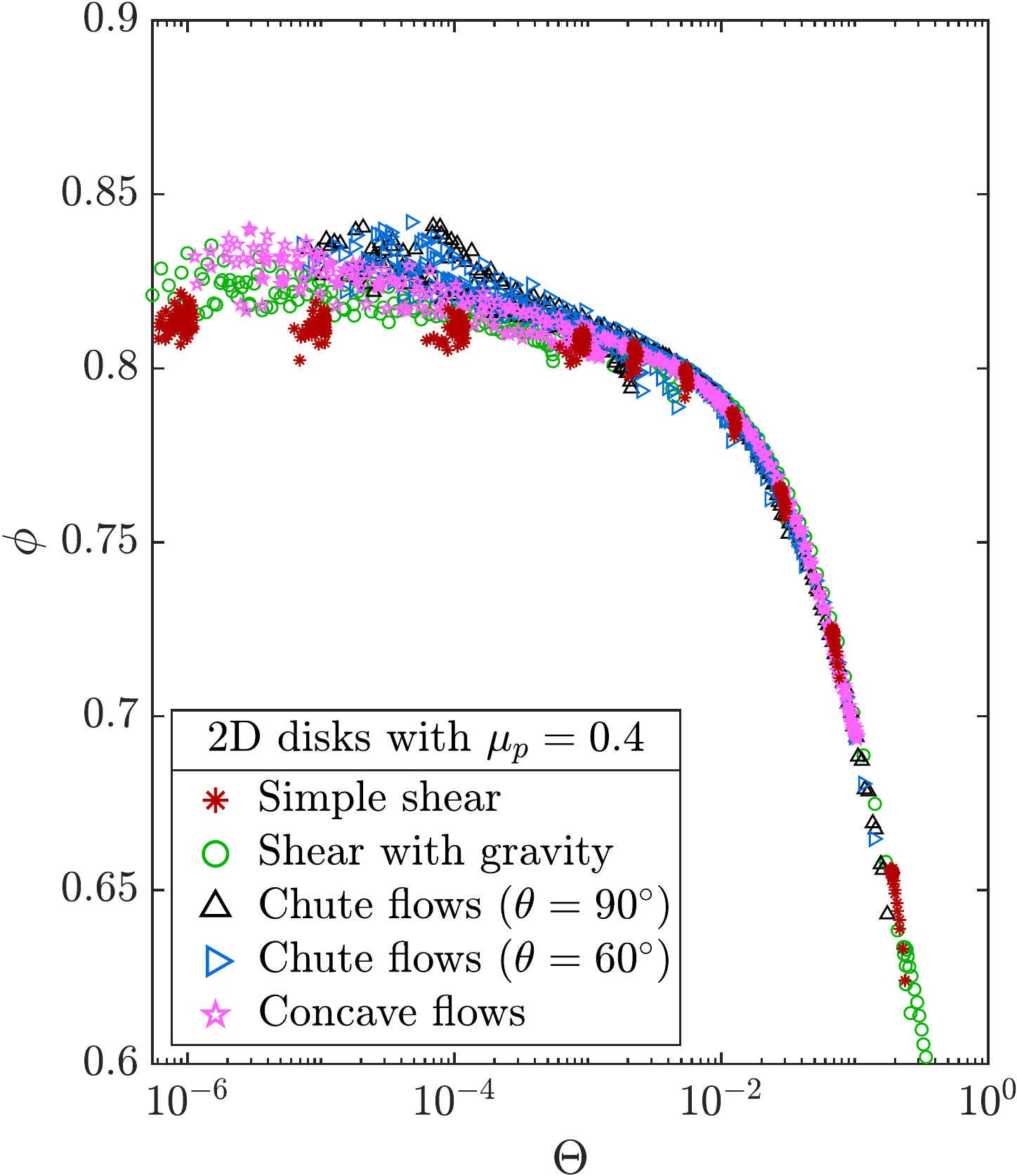}}
        \label{fig:S3c}
    \end{subfigure}
    \quad
    \begin{subfigure}[t]{0.232\textwidth}  
        \centering 
        \phantomcaption
        \stackinset{l}{0.84\textwidth}{b}{1.01\textwidth}
        {(\thesubfigure)}
        {\includegraphics[width=\textwidth]{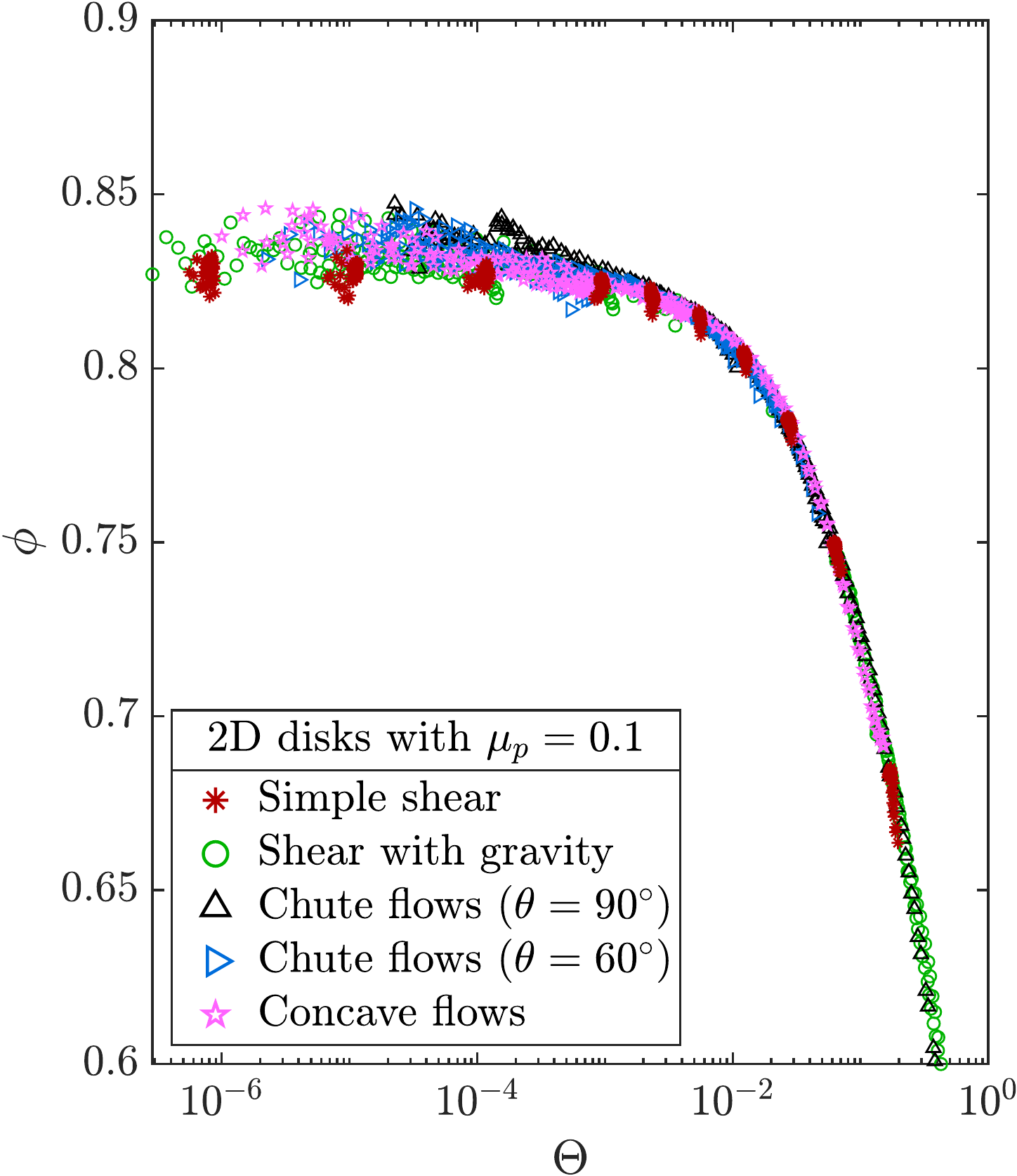}}
        \label{fig:S3d}
    \end{subfigure}
     \captionsetup{justification=raggedright, singlelinecheck=false}
    \caption{ DEM data from planar shear tests.
    Non-collapse of $\phi$ vs $\Theta$ for 3D spheres with $\mu_p=0.4$ (\protect\subref{fig:S3a}) and $\mu_p=0.1$ (\protect\subref{fig:S3b}), and 2D disks with $\mu_p=0.4$ (\protect\subref{fig:S3c}) and $\mu_p=0.1$ (\protect\subref{fig:S3d}).
    }
    \label{fig:S3}
\end{figure}

\begin{figure}[H]
    \centering
    \begin{subfigure}[t]{0.228\textwidth}
        \centering
        \phantomcaption
        \stackinset{l}{0.84\textwidth}{b}{1.03\textwidth}
        {(\thesubfigure)}
        {\includegraphics[width=\textwidth]{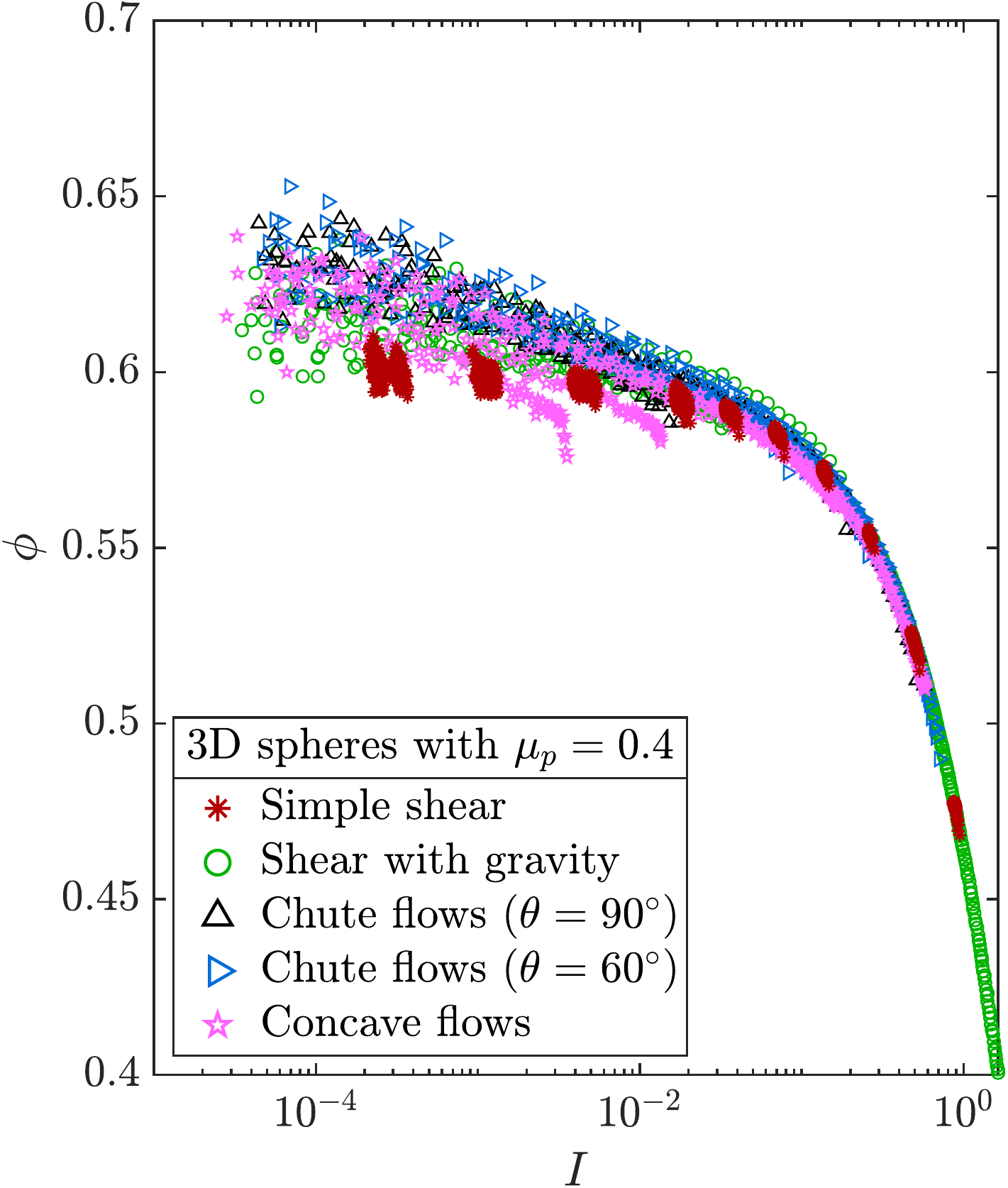}}
        \label{fig:S4a}
    \end{subfigure}
    \quad
    \begin{subfigure}[t]{0.228\textwidth}  
        \centering
        \phantomcaption
        \stackinset{l}{0.84\textwidth}{b}{1.03\textwidth}
        {(\thesubfigure)}
        {\includegraphics[width=\textwidth]{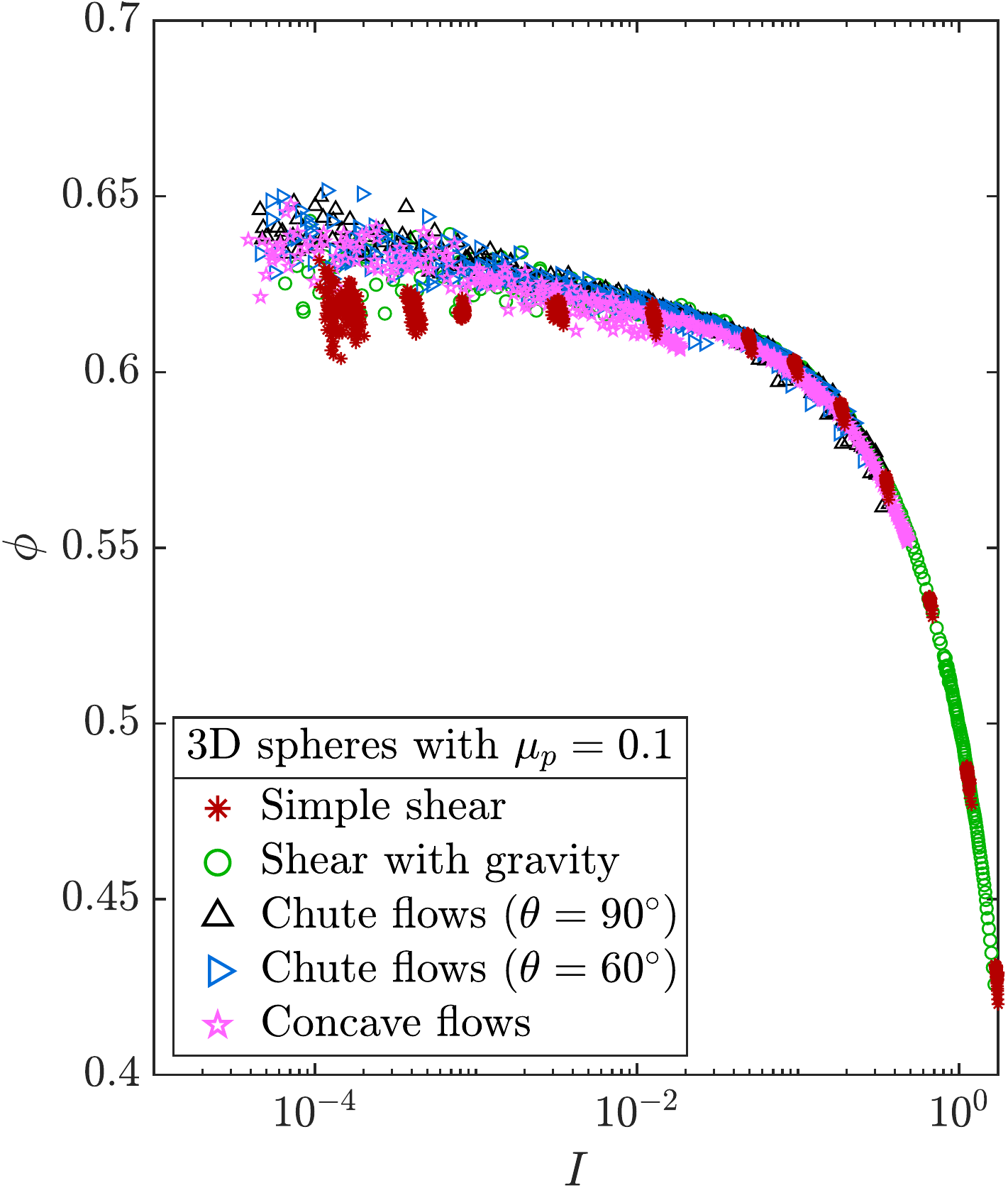}}
        \label{fig:S4b}
    \end{subfigure}
    \quad
    \begin{subfigure}[t]{0.232\textwidth}  
        \centering 
        \phantomcaption
        \stackinset{l}{0.84\textwidth}{b}{1.01\textwidth}
        {(\thesubfigure)}
        {\includegraphics[width=\textwidth]{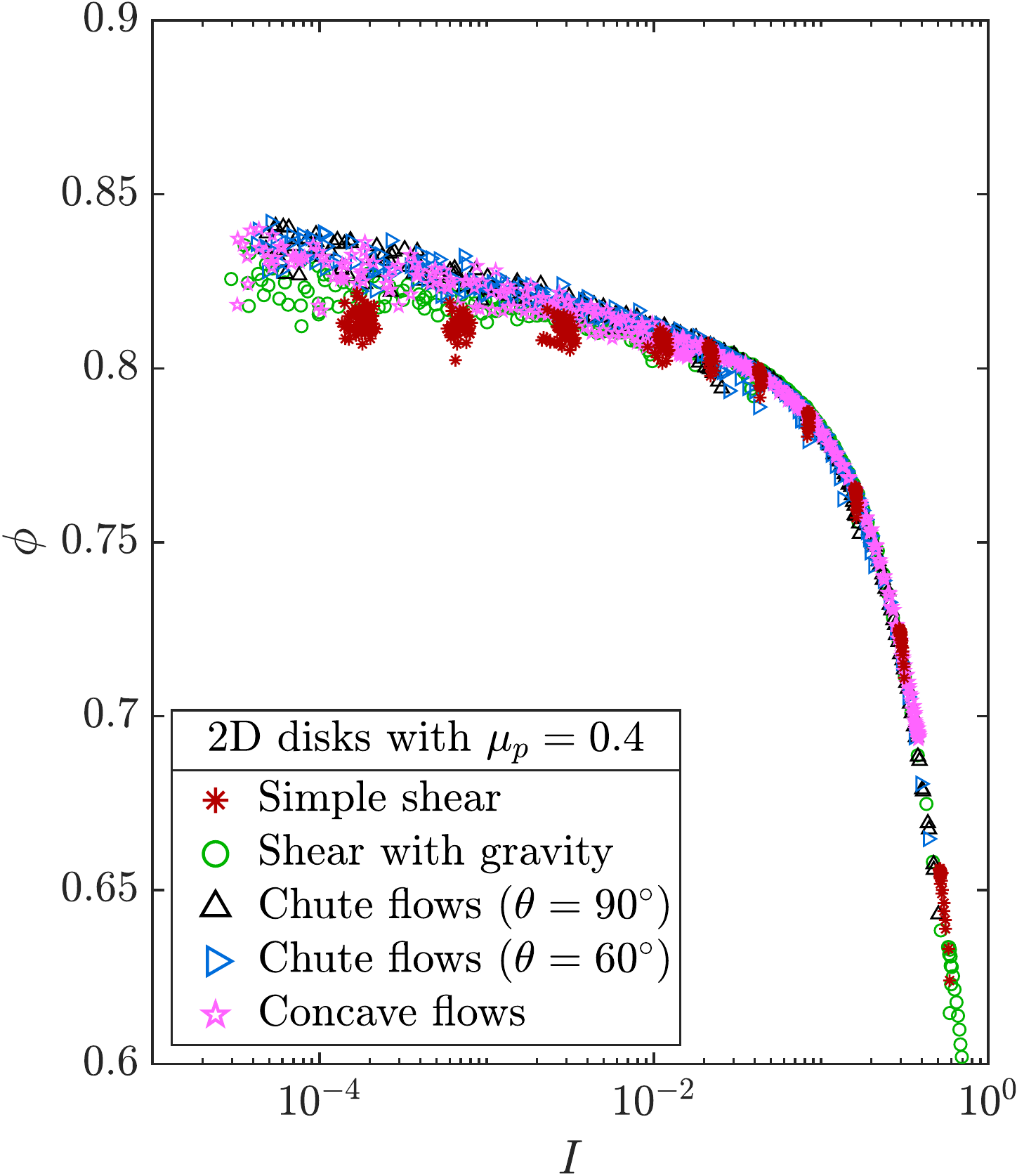}}
        \label{fig:S4c}
    \end{subfigure}
    \quad
    \begin{subfigure}[t]{0.232\textwidth}  
        \centering 
        \phantomcaption
        \stackinset{l}{0.84\textwidth}{b}{1.01\textwidth}
        {(\thesubfigure)}
        {\includegraphics[width=\textwidth]{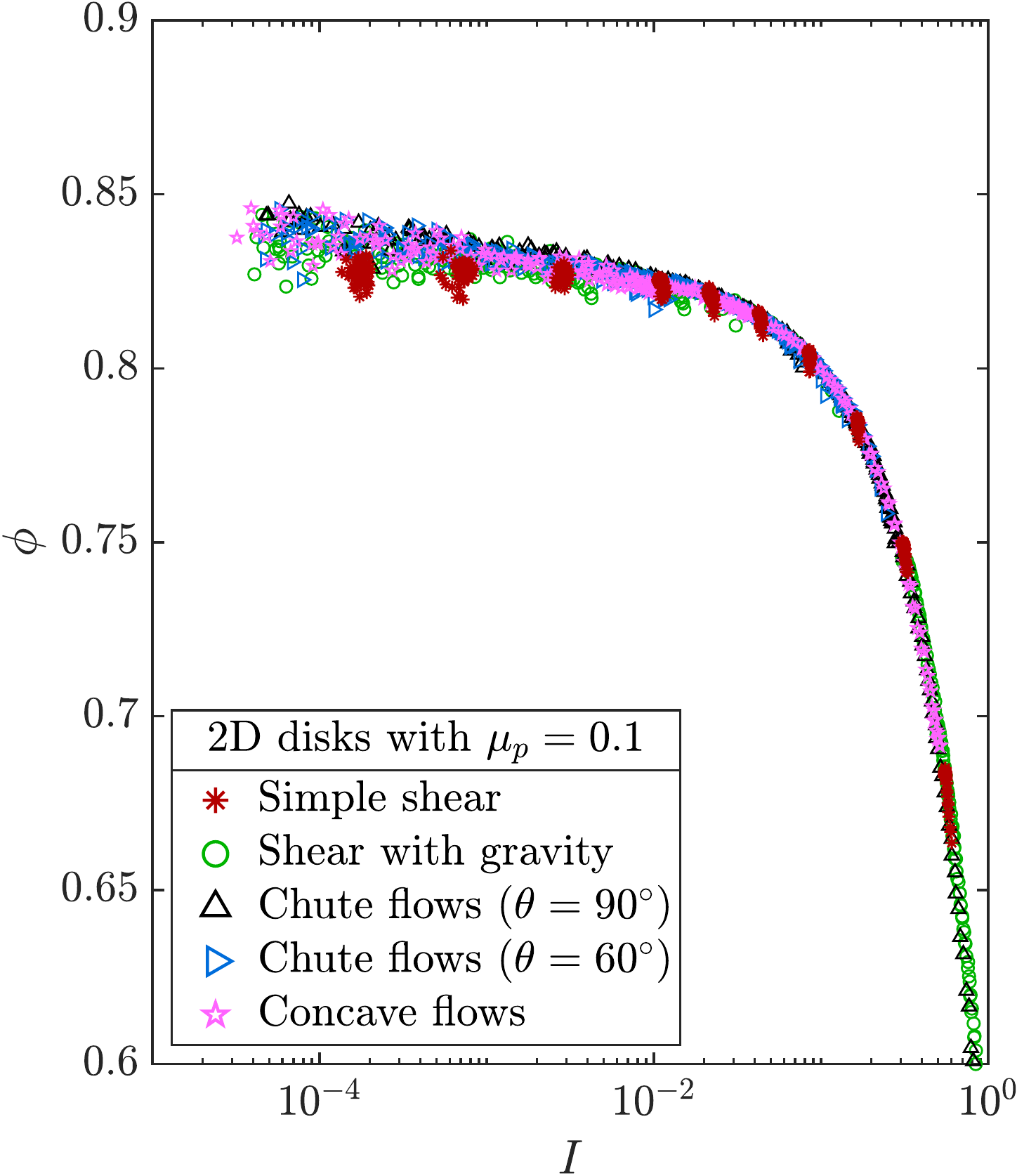}}
        \label{fig:S4d}
    \end{subfigure}
     \captionsetup{justification=raggedright, singlelinecheck=false}
    \caption{ DEM data from planar shear tests.
    Non-collapse of $\phi$ vs $I$ for 3D spheres with $\mu_p=0.4$ (\protect\subref{fig:S4a}) and $\mu_p=0.1$ (\protect\subref{fig:S4b}), and 2D disks with $\mu_p=0.4$ (\protect\subref{fig:S4c}) and $\mu_p=0.1$ (\protect\subref{fig:S4d}).
    }
    \label{fig:S4}
\end{figure}

\begin{figure}[H]
    \centering
    \begin{subfigure}[t]{0.27\textwidth}
        \centering
        \phantomcaption
        \stackinset{l}{-0.02\textwidth}{b}{1.02\textwidth}
        {(\thesubfigure)}
        {\includegraphics[width=\textwidth]{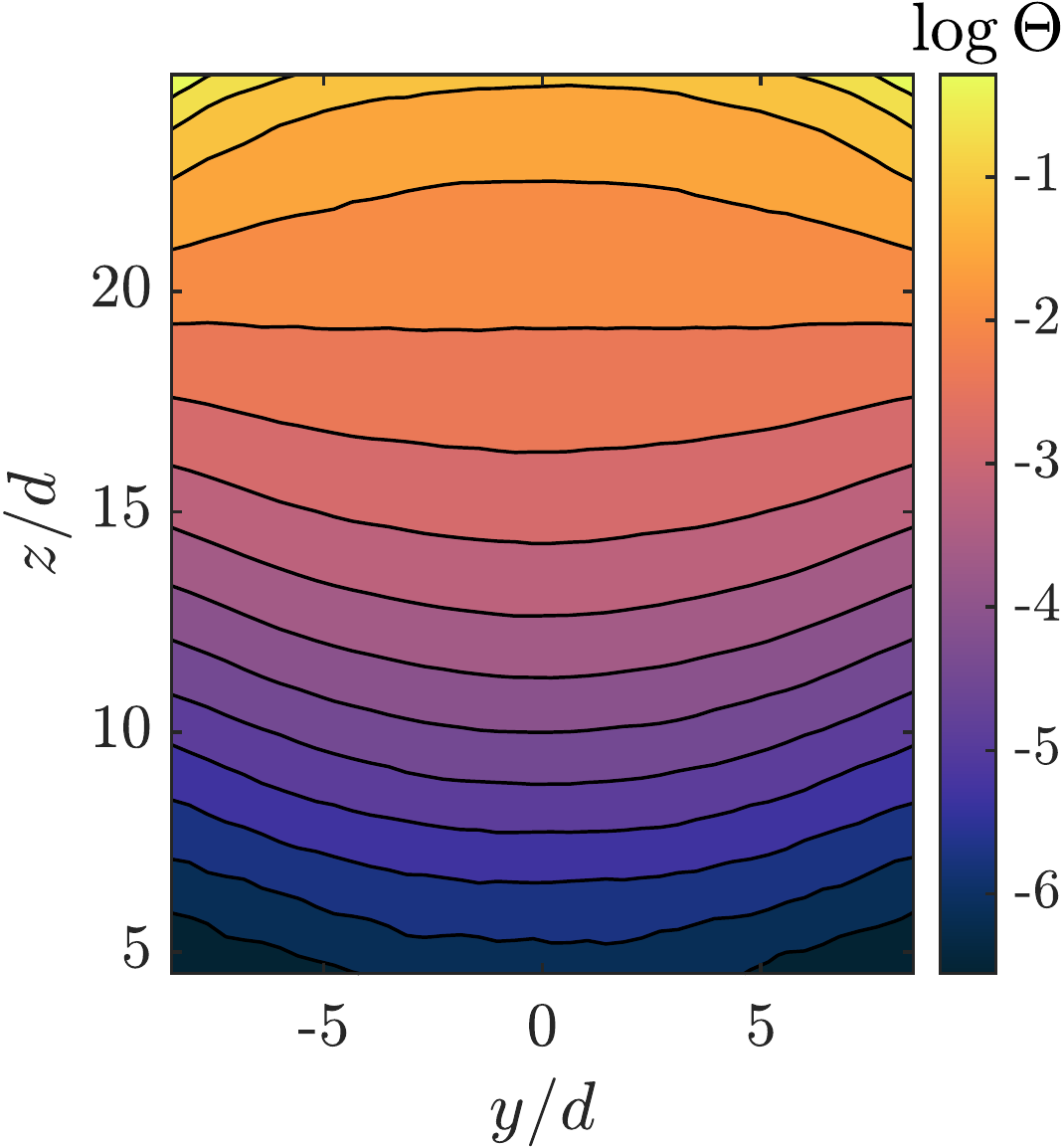}}
        \label{fig:S5a}
    \end{subfigure}
    \quad
    \quad
    \begin{subfigure}[t]{0.23\textwidth}  
        \centering
        \phantomcaption
        \stackinset{l}{-0.05\textwidth}{b}{1.21\textwidth}
        {(\thesubfigure)}
        {\includegraphics[width=\textwidth]{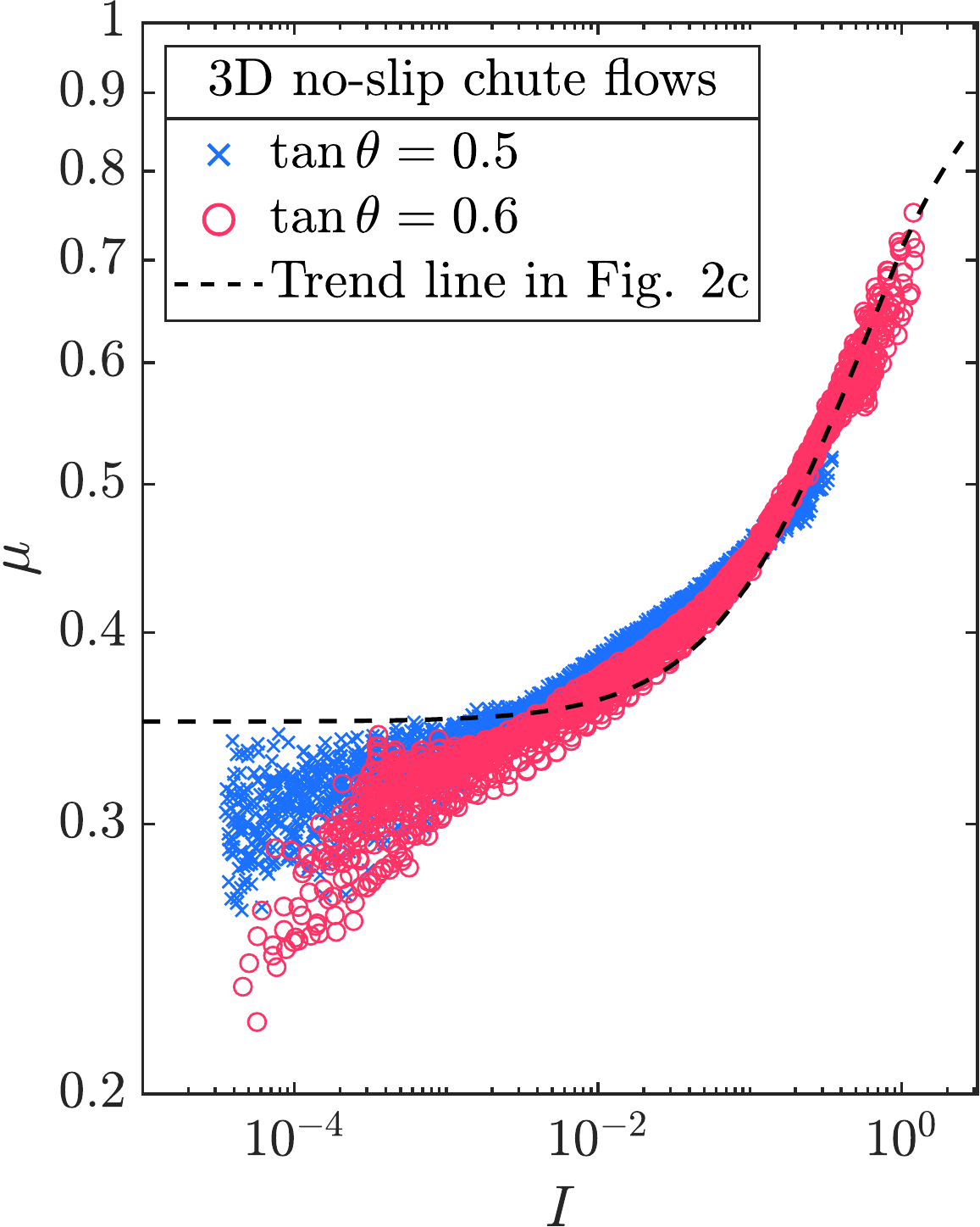}}
        \label{fig:S5b}
    \end{subfigure}
    \quad
    \quad
    \begin{subfigure}[t]{0.235\textwidth}  
        \centering 
        \phantomcaption
        \stackinset{l}{-0.01\textwidth}{b}{1.2\textwidth}
        {(\thesubfigure)}
        {\includegraphics[width=\textwidth]{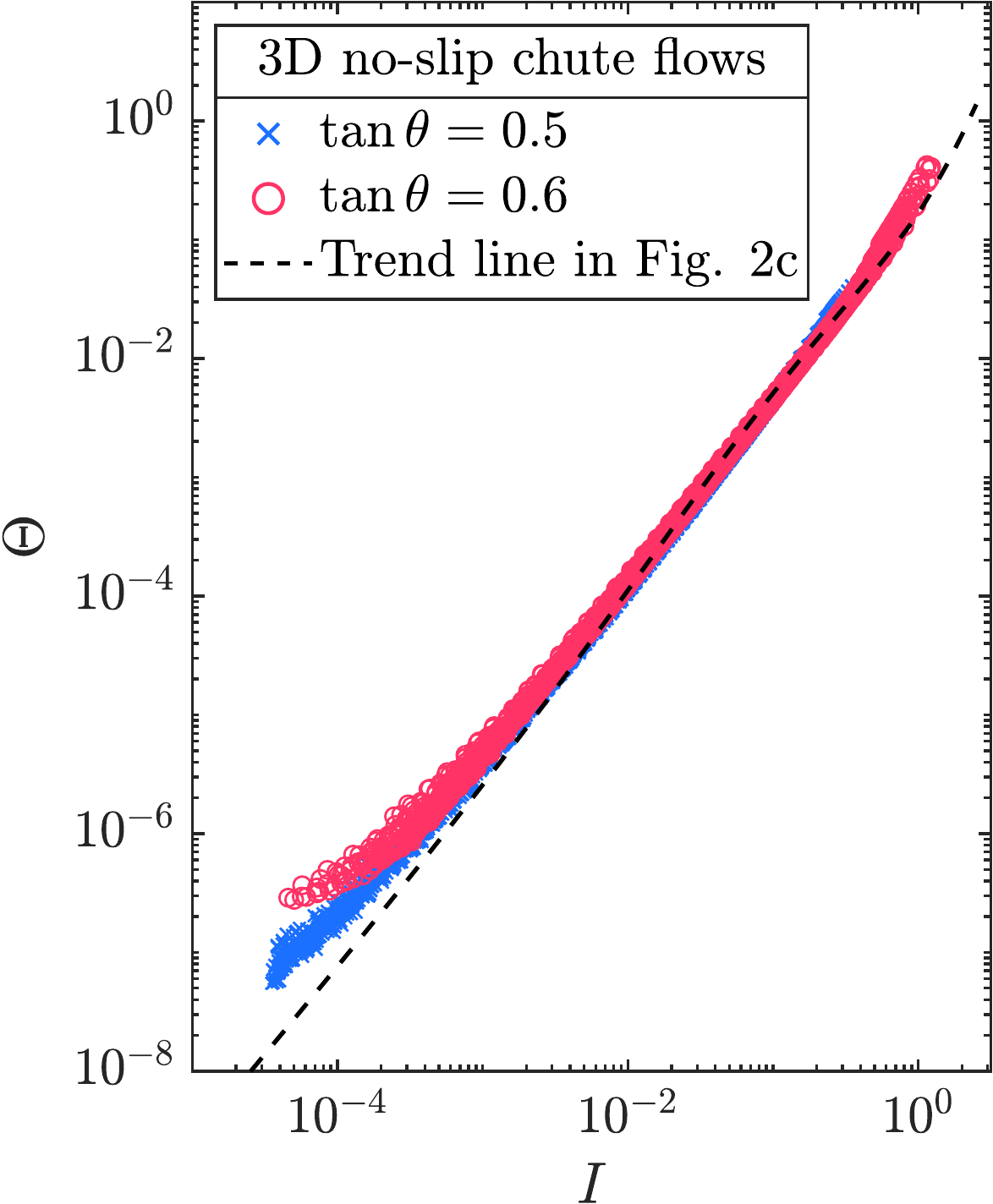}}
        \label{fig:S5c}
    \end{subfigure}
     \captionsetup{justification=raggedright, singlelinecheck=false}
    \caption{ DEM data from inclined chute flows with no-slip sides.
    (\protect\subref{fig:S5a}) The distribution of $\log{\Theta}$ for $\tan{\theta}=0.6$. (\protect\subref{fig:S5b}) Non-collapse of $\mu$ vs $I$ (\protect\subref{fig:S5b}) and $\Theta$ vs $I$ (\protect\subref{fig:S5c}). Trend lines are from the planar shear tests.
    }
    \label{fig:S5}
\end{figure}

\begin{figure}[H]
    \centering
    \begin{subfigure}[t]{0.255\textwidth}
        \centering
        \phantomcaption
        \stackinset{l}{-0.03\textwidth}{b}{1.25\textwidth}
        {(\thesubfigure)}
        {\includegraphics[width=\textwidth]{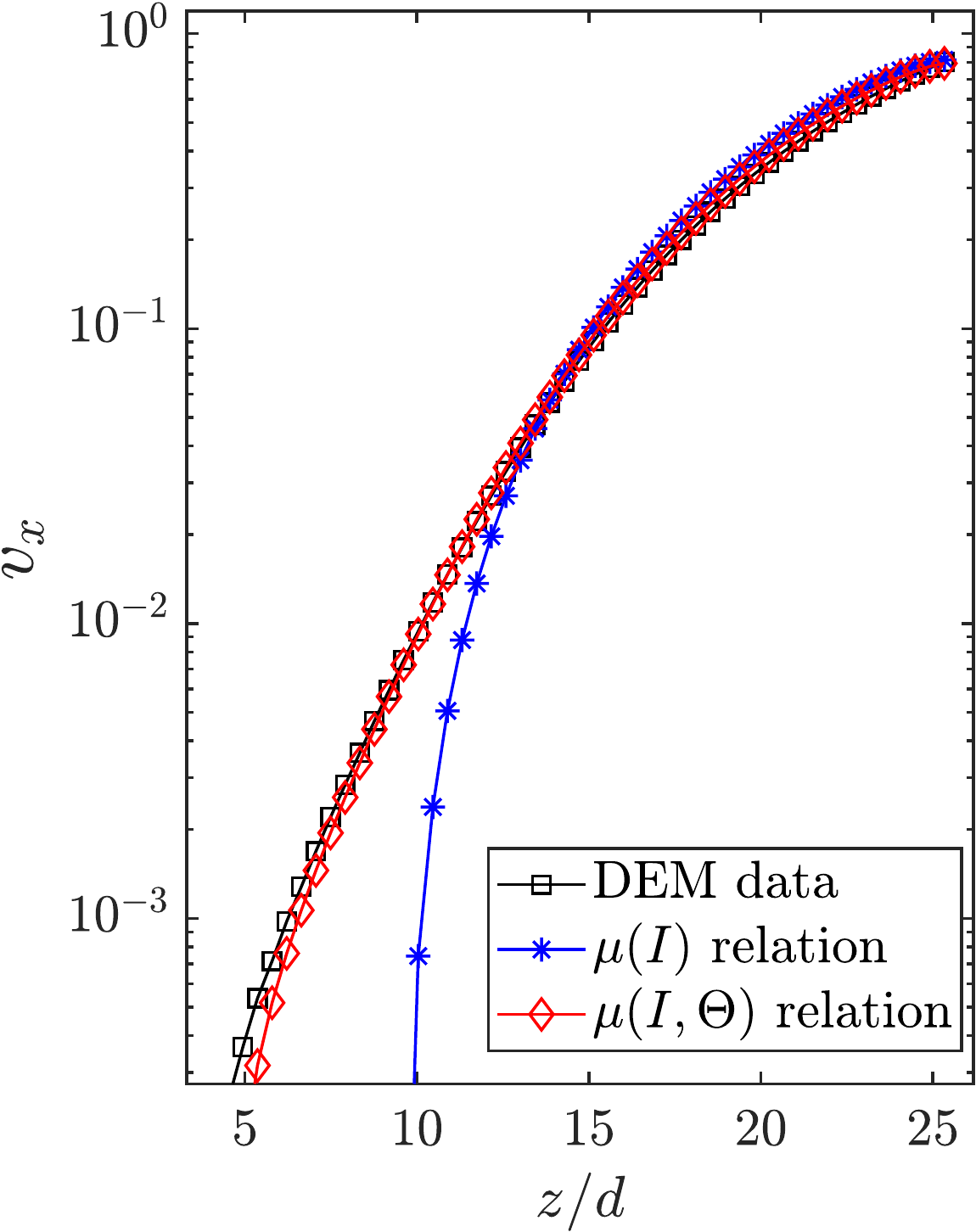}}
        \label{fig:S6a}
    \end{subfigure}
    \quad
    \quad
    \begin{subfigure}[t]{0.27\textwidth}  
        \centering
        \phantomcaption
        \stackinset{l}{-0.04\textwidth}{b}{1.19\textwidth}
        {(\thesubfigure)}
        {\includegraphics[width=\textwidth]{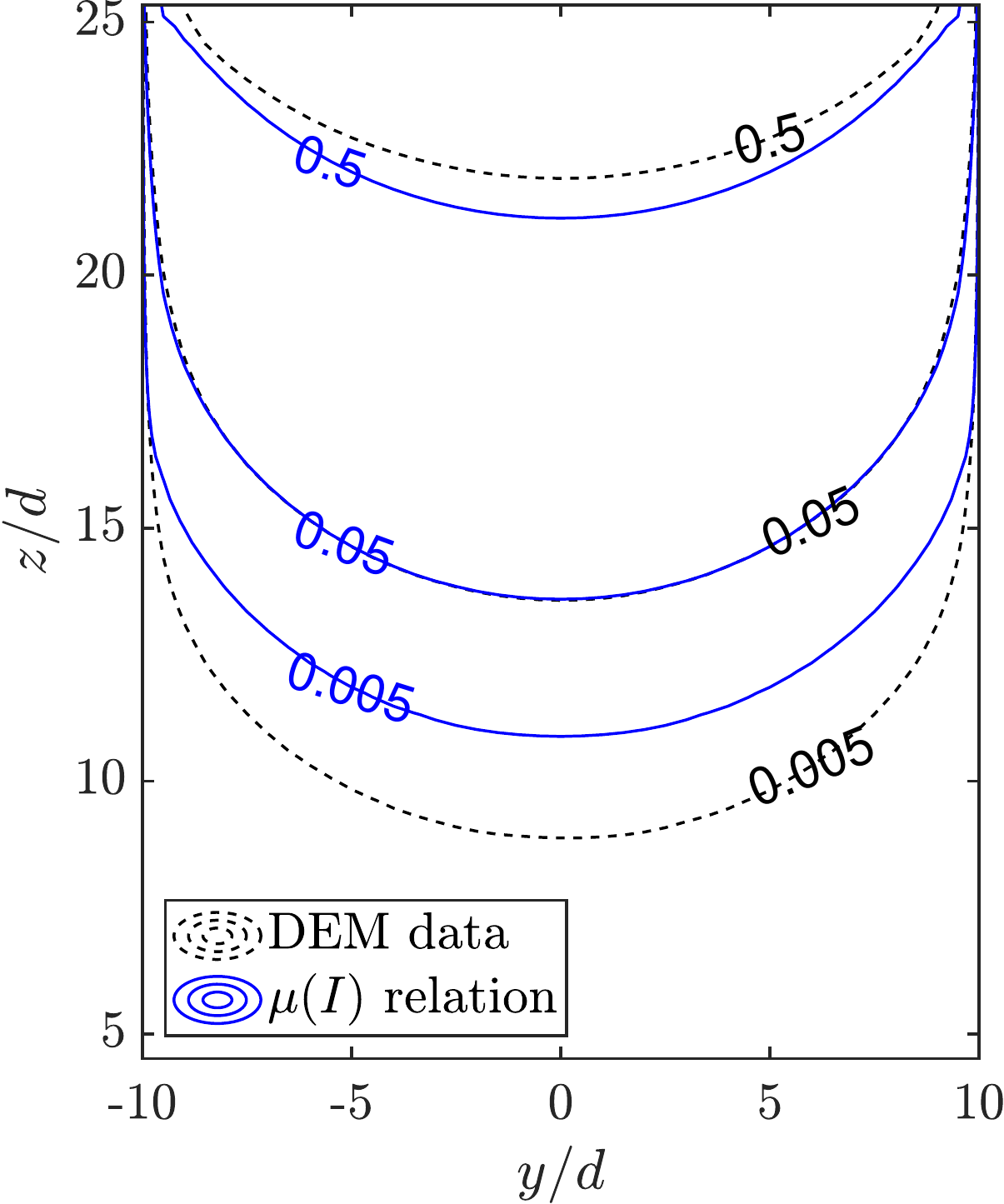}}
        \label{fig:S6b}
    \end{subfigure}
    \quad
    \quad
    \begin{subfigure}[t]{0.27\textwidth}  
        \centering 
        \phantomcaption
        \stackinset{l}{-0.01\textwidth}{b}{1.2\textwidth}
        {(\thesubfigure)}
        {\includegraphics[width=\textwidth]{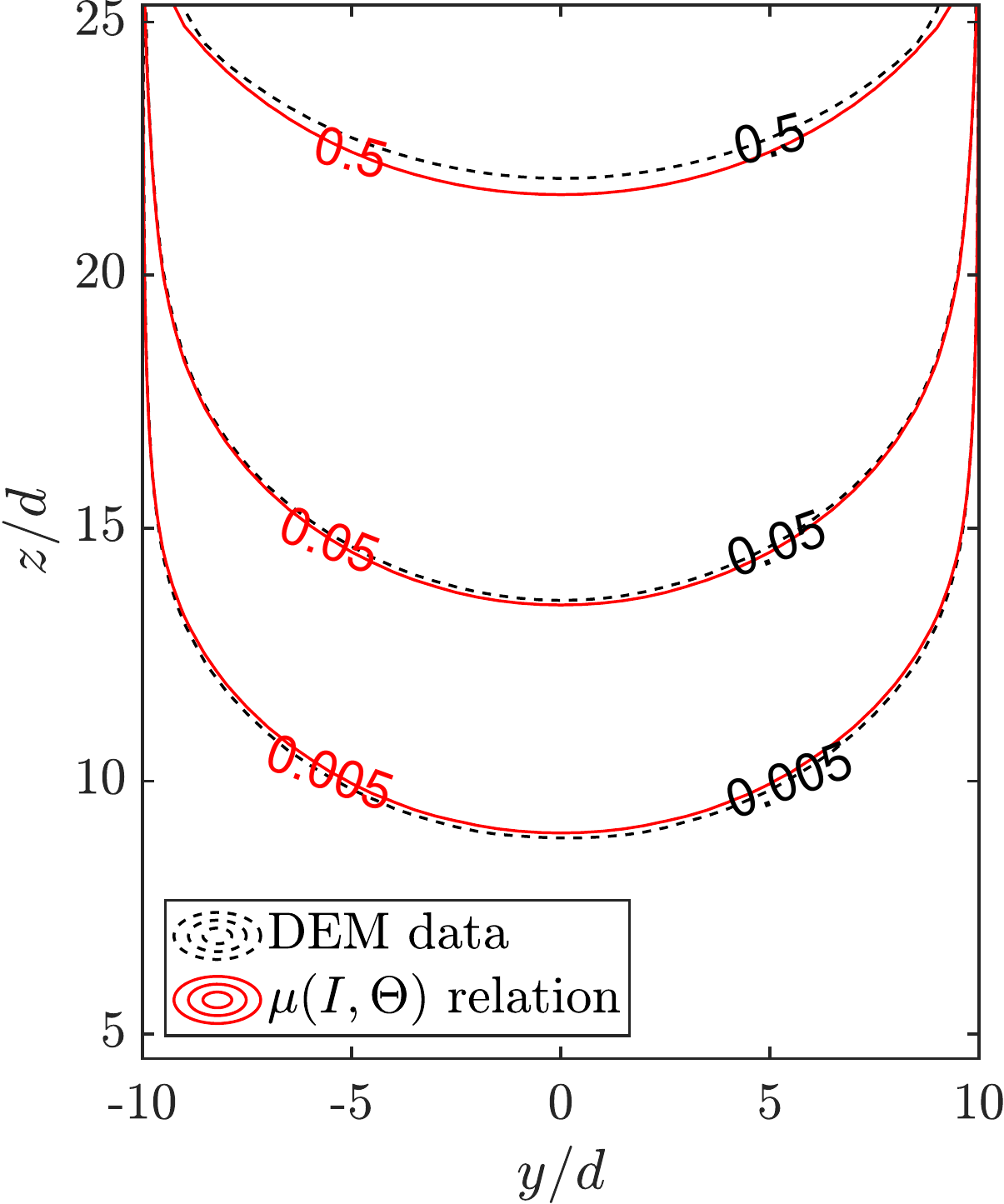}}
        \label{fig:S6c}
    \end{subfigure}
     \captionsetup{justification=raggedright, singlelinecheck=false}
    \caption{ Simultaneous solutions of the Cauchy momentum equation with the $\mu(I,\Theta)$ relation in inclined chute flows with no-slip sides for $\tan{\theta}=0.6$ ($\tan{\theta}=0.5$ case is shown in the main text). (\protect\subref{fig:S6a}) Comparing model velocity to DEM viewed down the chute's center-plane (fixed $y$).
    (\protect\subref{fig:S6b}) Comparing DEM velocity field to that obtained from solving the $\mu(I)$ relation and (\protect\subref{fig:S6c}) the $\mu(I, \Theta)$ relation.
    }
    \label{fig:S6}
\end{figure}

We provide additional figures from the DEM simulations. Fig.~\ref{fig:S1} to \ref{fig:S4} show the relations between $\mu$, $I$, $\Theta$, and $\phi$ obtained from the planar shear flows. Fig.~\ref{fig:S5} and Fig.~\ref{fig:S6} show supplemental DEM data and solutions to the Cauchy momentum equation in the inclined chute geometry.

\section{Continuum Simulation Method}
\begin{figure}[H]
    \centering
        \includegraphics[width=0.4\textwidth]{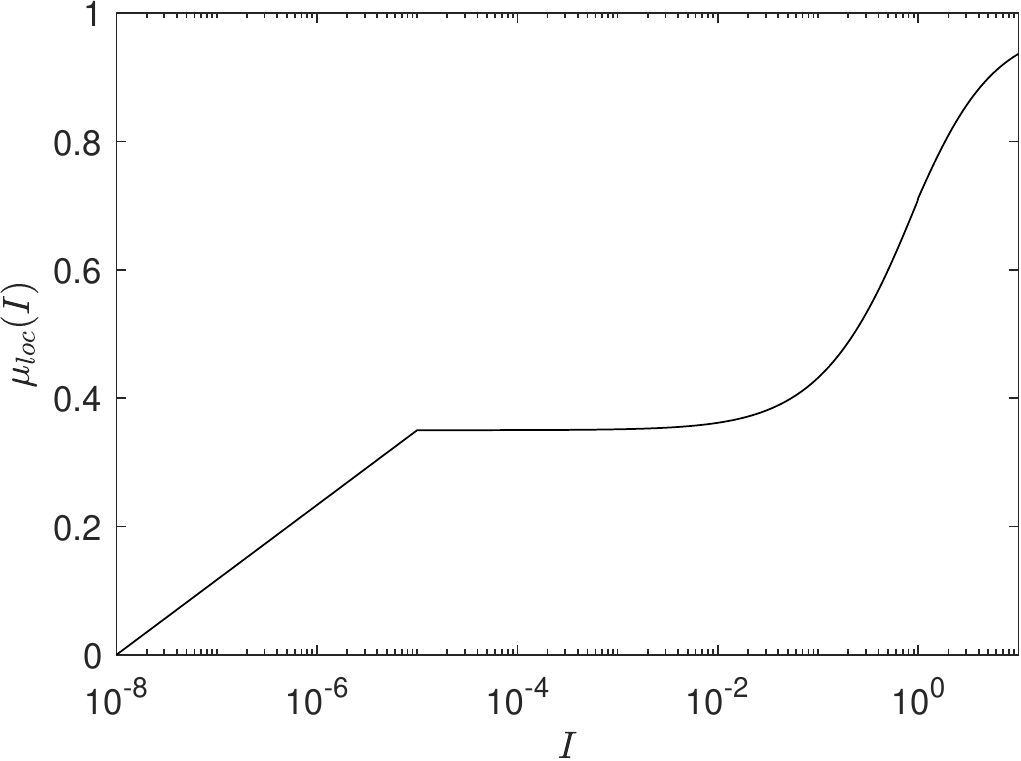}
        \caption{Modified $\mu_{loc}(I)$ for regularization.}
        \label{fig:S7}
\end{figure}

We use the finite difference method to solve the Cauchy momentum equation in the inclined chute flows. The velocity field is calculated on a $41\times50$ grid representing the $yz$ plane. The stress field is staggered, located on cell centers (a $40\times 49$ grid of locations). $\Theta$ is interpolated to the grid of stress. For regularization, $\mu_{loc}(I)$ is modified to gradually vanish from $I=10^{-5}$ to $I=10^{-8}$ which prevents numerical errors by giving a finite but insignificant shear rate for $\mu<\mu_s$ (Fig.~\ref{fig:S7}). We assume that the stress deviator $\boldsymbol{\sigma}'$ and the strain-rate tensor $\bf{D}$ are co-directional: $\bf{{D}/{|D|}}={\boldsymbol{\sigma'}/{|\boldsymbol{\sigma'}|}}$.
We apply $\sigma_{xz},\sigma_{yz}=0$ at the surface by assuming the surface is flat and imposing an imaginary stress of opposite sign mirrored across the surface. We know analytically that with co-directional flow rules, steady flows always develop lithostatic pressure, which we exploit by pre-setting $\sigma_{zz}=\rho_s\phi G\cos\theta(H-z)$. We update the velocity field putting either $\mu=\mu_{loc}(I)$ or $\mu=\Theta^{-1/6}f(I)$ in the momentum equation until the Frobenius norm of the velocity change becomes small enough. We have checked that the final results are independent of the initial velocity.

\section{Video}
Download ``Inclined\_Chute\_Flows\_tan05.avi'' to watch the motion of particles in the inclined chute flow with $\tan{\theta}=0.5$ including the part flowing in the opposite direction. The middle part receives a gravitational acceleration of $\vec{G}=G\sin{\theta}\hat{x}-G\cos{\theta}\hat{z}$, while the other half receives $\vec{G}=-G\sin{\theta}\hat{x}-G\cos{\theta}\hat{z}$, which naturally sets the average velocity to vanish at the boundaries.

\end{document}